# Progress in Developing Highly Efficient *p*-Type TCOs for Transparent Electronics: A Comprehensive Review

Jarnail Singh[1] and Pankaj Bhardwaj[2,*]

[1]Department of Mechanical Engineering, University Centre for Research and Development, Chandigarh University, Mohali 140413, Punjab, India

[2]Department of Material Science and Engineering, National Institute of Technology Hamirpur-177005

[*]*E-mail address:* pjbh27@gmail.com

***ORCID ID:*** *0000-0001-8341-525X*

**Abstract**

Transparent conducting oxides (TCOs) represent a remarkable class of materials that possess both excellent electrical conductivity and high optical transparency, which are typically considered mutually exclusive in traditional materials. This unique combination of properties makes TCOs highly desirable for various applications, particularly in optoelectronic devices such as transparent electrodes for displays, solar cells, and touchscreens. However, developing high-performance TCOs, especially *p*-type materials, has been a significant challenge. In conventional materials, achieving both high electrical conductivity and optical transparency is difficult. Because materials with wide optical band gap (≥ 3.1 eV) are transparent in the visible region but lack electrical conductivity, while conductive metals are opaque. Hence, the only way to induce both properties in a single material is to create non-stoichiometry and/or defects. By introducing shallow defects near the conduction band for *n*-type materials and the valence band for *p*-type materials, it is possible to enhance the conductivity of the material at room temperature. However, developing efficient *p*-type TCOs has been particularly difficult due to the localized nature of the valence band derived from O 2p orbitals and the challenges associated with shallow acceptors, resulting in large effective mass of holes. While commercially available TCOs are predominantly *n*-type, such as Sn:$In_2O_3$, Al:ZnO, and F:$SnO_2$, the development of efficient *p*-type TCOs lags. In this review, we have discussed the origin of *p*-type conductivity in TCOs, and the difficulties encountered in developing efficient *p*-type materials. We have also demonstrated the fundamental materials physics of *p*-type TCOs, including electronic structures, doping, defect properties, and optical properties. A range of deposition techniques has been adopted to prepare

TCO films, and this review provides a detailed discussion of these techniques and their relative deposition parameters. Overall, we have presented an up-to-date and comprehensive review of different *p*-type transparent conducting oxide thin films, providing insights into ongoing research and potential future directions in this field.

## 1. Introduction

Transparent conducting oxide (TCO) materials hold excellent potential in a wide variety of applications, including thin-film solar cells, flat panel display, liquid crystal display (LCDs), light-emitting diodes (LEDs), thin-film transistors (TFTs), smart windows, invisible circuits, flexible and transparent devices [1]. For instance, transparent display devices with great refresh frequency and excellent resolution or pixel density can be realized by employing highly effective TCO materials. Furthermore, energy-efficient displays that operate in typical lighting atmospheres can be fabricated using TCO materials. TCOs may allow the removal of panel backlight (that accounts for power consumption around 90% in current displays) [2,3]. However, TCOs acquire a wide range of optoelectronic applications, but very modest work has been done on the development of *p*-type TCOs. The *p*-type TCOs with comparable performance to their *n*-type counterparts are necessary for the fabrication of active electronic devices based on transparent *p*-*n* junction. But the *p*-type TCOs were missing in literature for a long time until Kawazoe and co-workers (from Tokyo Institute of Technology, Japan) in 1997 came up with an article in which *p*-type conductivity is reported in the thin film form of $CuAlO_2$ [4]. This discovery helped other researchers to realize that the development of *p*-type TCOs is not an unfeasible task and unlocked a new field of **"invisible electronics"** or **"transparent electronics"**.

However, it is worth discussing that the first *p*-type semi-transparent conducting material in the form of nickel oxide (NiO) was reported by Sato et al. in 1993 [5]. They have deposited a thin film of NiO at various $O_2$ content by using RF magnetron sputtering technique. A very low value of resistivity of 0.14 Ω-cm with a carrier concentration of $1.3\times10^{19}$ $cm^{-3}$ was observed in pristine NiO thin film at substrate temperature and oxygen partial pressure of 200 $^{o}$C and 1.33 Pa, respectively. Hall measurement results demonstrated the *p*-type conductivity in NiO thin film. But transparency was not upto the mark with an average value of 40% in the visible region for 110 nm thin film of NiO. They have also fabricated the p-i-n diode on a glass substrate, consisting of *p*-NiO/i-NiO/i-ZnO/*n*-ZnO and observed that optical transparency further narrows down to 20%, which cannot be considered for superior optoelectronics devices-based applications. Still, this report has turned out to be a landmark in the field of **“invisible electronics”** and optoelectronic technology development.

The research for developing *p*-type TCOs has been accelerated when Kawazoe et al. demonstrated the significant hole conductivity at room temperature [4]. Copper aluminium oxide ($CuAlO_2$) thin film having a thickness of 500 nm was deposited on (0001) sapphire substrate using pulsed laser deposition (PLD) technique. As-deposited $CuAlO_2$ thin film illustrated a wide optical band of 3.5 eV. The presence of $Al^{3+}$ ions in between $Cu^+$ ions reduced the cross-linking of $d^{10}$ electrons of $Cu^+$ network along the c-axis. This may be responsible for the larger optical band gap of $CuAlO_2$. The electrical conductivity of 1 S/cm was achieved in $CuAlO_2$ thin film. The positive value of Hall coefficient and Seebeck coefficient confirmed the *p*-type nature of $CuAlO_2$ thin film. Excellent Hall mobility of 10.4 $cm^2/Vs$ (like that of ITO) is responsible for the significant conductivity value of *p*-type $CuAlO_2$. However, its conductivity value is still lower than the highly effective *n*-type TCOs. Nevertheless, these results demonstrated that both properties' simultaneous existence in *p*-type TCOs is not an invincible goal. After this report, intensive research has been conducted to explore other materials as *p*-type TCOs. Here, we have examined the fundamental physics of TCOs and performed a comprehensive material-wise literature survey of *p*-type transparent conductors, including synthesization methods, deposition techniques, and their optoelectronic properties.

## 1.1 Fundamental Material Physics of TCOs

### 1.1.1 Electrical properties

The simultaneous existence of excellent electrical conductivity alongwith high optical transparency in stoichiometric oxide materials is not feasible. The electrical conductivity mechanism in TCO material can be explained based on classical mechanics on charge transport in metals, i.e., Drude theory. The free charge carriers generated due to doping can be considered as free-electron gas. Hence, their conductivity is directly related to the density of charge carriers and their effective mass:

$$\sigma = n_e e \mu_e = \frac{n_e e^2 \tau}{m_e} \tag{1}$$

where σ is the electrical conductivity, $n_e$ is the number of charge carriers, $m_e$ is the effective mass of charge carriers, $\mu_e$ is the mobility of charge carriers, e is the electronic charge, and τ is relaxation time. The conduction band cannot be thermally populated at room temperature due to the wide optical band of TCO materials. Hence, stoichiometric TCO materials are insulating in nature. In the case of intrinsic TCO materials, availability of conduction electrons or holes is mainly due to unintentional defects such as cation interstitials or anion vacancies (in case of *n*-type) and cation vacancies or anion interstitials (in case of *p*-type). This type of defects may create shallow donor and acceptor levels close to conduction and valence band, respectively (as shown in figure 1). And these excess generated charge carriers can be thermally ionized at room temperature and impart conduction within the material. To further enhance its electrical properties, extrinsic doping can be introduced into TCO materials.

Hence, the basic idea in TCO materials is to generate high mobile free charge carriers (whether it is electrons or holes) into wide optical band gap semiconductors/insulators by using appropriate dopants. Considering the example of highly effective *n*-type TCO, i.e., Indium-tin oxide (ITO), the substitution of $Sn^{4+}$ cations at the $In^{3+}$ cationic sites lead to the formation of donor-defects below the conduction band minimum (CBM). Apart from that, doping of Sn into $In_2O_3$ material also leads to the dispersion of conduction band, which can reduce the effective mass of charge carriers and impart conductivity to the material. Therefore, Sn-doping into $In_2O_3$ material results in high electrical conductivity of the order of $10^4$ S/cm at room temperature with minimal effect on its optical transparency [6]. Hence, TCO material exhibits excellent electrical conductivity alongwith optical transparency in the visible region.

One thing common in highly effective TCO materials (such as ITO, ZnO, $SnO_2$, and CdO) is their similar structural, chemical, and electronic properties. For instance, post-transition metal oxides have $(n\text{-}1)d^{10}ns^2$ outer shell orbitals with closely packed dense structures. There is a strong interaction between metal ns orbital and oxygen 2p orbital. The conduction band is formed by the hybridization of anti-bonding metal ns orbitals and oxygen p orbitals. In contrast, the valence band is shaped by interaction between bonding and non-bonding oxygen 2p states. Interaction between metal ns and oxygen 2p orbitals leads to the forbidden gap between the valence band and conduction band. As s-orbital is highly delocalized in spatially spread, it leads to the large dispersion of conduction band by overlapping with oxygen 2p orbitals. Besides, cations having octahedral coordination with oxygen anions results in further dispersion of conduction band. Considering the case of rocksalt CdO, large dispersion is observed due to octahedral coordination with oxygen anions, leading to the lowest effective mass of electrons [7].

Hence, all highly effective TCOs demonstrate excellent electrical conductivity at room temperature due to the low effective mass of charge carriers. Their electrical properties can further be modified by introducing appropriate counter cations or anions. Replacement of host cation with suitable aliovalent cation can result in the generation of free charge carriers, which can impart conductivity to the material. Most highly effective TCOs doped with suitable countercation (without affecting its crystal structure) lead to remarkable optoelectronic properties in a single material.

Besides, removing or replacing oxygen anions from its lattice sites can also provide more electrons within the crystal system, in the case of *n*-type TCO material. On the contrary, in the case of *p*-type TCOs, incorporation of oxygen at the interstitial positions generates excess charge carriers (holes) into the lattice [8]. These generated charge carriers may be free or localized around the vacant site, ascribed to the formation energy of oxide material. Conventional TCO materials have low formation energy, exhibit a higher density of free charge carriers. On the other hand, most of the *p*-type TCOs are attributed to higher formation energy. Hence, the charge carrier generated localized around oxygen 2p states, which leads to the formation of a deep acceptor level [9].

### 1.1.2 Optical properties

Maxwell's equation of electromagnetic (EM) waves demonstrated that electrical properties and optical properties are entirely contradictory to each other. When EM waves traveled through a semiconducting material (uncharged), the solution to Maxwell's equation determines the value of the refractive index in the form of real and imaginary parts, given below [10,11]:

$$n^2 = \frac{\varepsilon}{2}\left[\left\{1+\left(\frac{2\sigma}{\nu}\right)^2\right\}^{1/2} +1\right] \tag{2}$$

$$k^2 = \frac{\varepsilon}{2}\left[\left\{1+\left(\frac{2\sigma}{\nu}\right)^2\right\}^{1/2} -1\right] \tag{3}$$

where, n is refractive index of material, $\epsilon$ is dielectric constant, $\sigma$ is electrical conductivity, k is extinction coefficient, and $\nu$ is the frequency of EM waves. Considering the case of insulating material, where electrical conductivity ($\sigma$) $\rightarrow$ 0, refractive index (n) $\rightarrow \epsilon^{1/2}$ and then, extinction coefficient (k) $\rightarrow$ 0. It concludes that insulating material is transparent to EM radiations. On the other hand, in perfect conductors, Maxwell's equation solution for transmitted and reflected components of the electric field vector is $I_T = 0$ and $E_I = - E_R$. Hence, excellent conductors are opaque to EM radiations. When EM waves incident on the surface of a perfect conductor, it gets reflected with 180$^o$ out-of-phase. So, from the solution of Maxwell's equation, it is concluded that insulators are transparent to EM waves, whereas conductors are not.

Also, Drude's model (free electrons in metals) explained the absorption of TCO material in the infra-red (IR) region [12–14]. The equation for the dielectric constant of a free electron gas can be written as:

$$\varepsilon(\omega) = \frac{1 - \frac{4\pi n e^2}{m}}{\omega^2 + i\frac{\omega}{\tau}} \tag{4}$$

where, $\epsilon$ is dielectric constant, $\tau$ is relaxation time, and $\omega$ is frequency of EM waves. From Drude's theory, it is revealed that EM waves cannot travel through a material with a negative value of dielectric constant because its wave vector is imaginary and decays exponentially. The EM waves incident on that material gets reflected. The plasma frequency or cut-off frequency can be determined by:

$$\omega_p = \left(\frac{ne^2}{m^* \varepsilon_o}\right)^{1/2} \tag{5}$$

where, $\omega_p$ is a plasma or cut-off frequency, $\epsilon_o$ is permittivity of free space, n is the number density of charge carriers, and $m^*$ is its effective mass. The material is transparent to those EM radiations whose frequency is greater than plasma frequency. Hence, TCO materials are transparent in visible and near infra-red (NIR) regions and reflect IR radiation (below plasma frequency), as shown in figure 2. The material is transparent in the range between the shorter wavelength band gap ($\lambda_g$) to the higher plasma wavelength ($\lambda_p$). Plasmon frequency and mobility are governed by the effective mass of charge carriers and optical band gap. When the energy of incident photons is higher than that of band gap of material, it is being absorbed by the material, while if it is less than the plasmon frequency, then it is being reflected. Because at a frequency higher than plasma frequency, electrons did not react to EM radiations with a varying electric field and hence, act as transparent dielectrics. Most of the TCO materials exhibit plasma frequencies in the NIR region and accordingly, render high optical transparency in the visible part. The plasma frequency is directly proportional to the concentration of free electrons. The plasma frequency and carrier concentration vary similarly to that of resistivity values. Therefore, by tailoring the plasmon frequency, the infra-red region can be blocked or transmitted by TCO materials [15].

When a material is degenerately doped with suitable counter cation, excess generated charge carriers occupy the lower energy level of conduction band and push the Fermi level to a higher energy state. In this case of degenerate doping, an electron can only be excited to a higher energy level if it has sufficient energy to surpass the bottom filled energy levels of occupied conduction band and reside in unoccupied conduction band energy level. It results in the rise of band gap of material, which is advantageous for its optical properties. This shift in optical band energy is known as Burstein-Moss shift or BM shift and can be determined by:

$$\Delta E_{BM} = \frac{\hbar^2}{2m^*}\left(3\pi^2 n\right)^{2/3} \tag{6}$$

where, $\Delta E_{BM}$ is magnitude of BM-shift, n is carrier concentration, and $m^*$ is effective mass. If transparent material is heavily doped and concentration of charge carriers is high enough, then the plasma edge will shift towards the red region of visible spectrum.

Alongwith that, when the bottom energy states of conduction band are completely occupied, then there is a probability that electrons at Fermi level ($E_{FL}$) may have sufficient energy to jump into the adjacent unoccupied energy band (next to above the conduction band). This difference in the two lowest conduction bands is known as 'secondary gap" ($E_{sg}$), as shown in figure 3. Hence, a high secondary band gap (must be greater than 3.1 eV) is also necessary for TCO materials. Because material with a large value of secondary band gap ($E_{sg}$) can forbid the intra-band transition and illustrates high optical transparency. A similar discussion can equally apply for *p*-type TCO material, where holes rather than electrons cause conduction.

Therefore, it is concluded that physical properties of TCO materials are interconnected, and it is very improbable that a single material can pull off all the requirements. Contradictions will always be there for some points to improve the physical properties of TCO materials. But the most important property of materials is their charge carrier type, as it is hard to modify. The basic functionalities can be determined from a device having a single charge carrier type material. Whereas most electronic devices require manipulation of both charge carrier types (i.e., *p*-type and *n*-type) to function.

### 1.1.3 Optical and electrical performance or Figure of merit (FOM)

Better performance TCOs comprise significant optical and electrical properties, which are inversely related to each other. Hence, researchers have developed a reliable parameter, i.e., figure of merit (FOM), which is used for assessing the performance of TCO materials. FOM value can be determined by using the following equation:

$$FOM = \frac{\sigma}{\alpha} = \frac{-1}{R_{sh} \ln T} \tag{7}$$

where, σ is electrical conductivity and α is absorption coefficient of TCO material. The term $R_s$ is the sheet-resistance (thickness-independent parameter) of TCO material's thin-film and T is the transmittance in visible region. FOM value is also used to quantify the performance of different TCO materials, as it provides a single number by combining both properties in a particular combination. And that value can be set as a benchmark for comparison of all TCO materials.

However, several issues need to be taken care of while considering the FOM value as a performance parameter. In research articles, maximum transmittance peak values are used rather than average transmittance values over the visible range (1.6–3 eV) for solar cells and displays applications. Thus, various low optical band gap materials with a high FOM value have been reported, regardless of acquiring limited optical transparency in the visible region. Another issue is that two materials can have the same FOM value despite having different optoelectronic properties. For example, one material with optical transparency of 20% and another material exhibiting optical transparency of 80% can have the same FOM value because the former's conductivity is higher enough. This amount of optical transparency (20%) is not acceptable for optoelectronic device-based applications. Hence, FOM values can be compared for those materials that exhibit considerable optical transparency in visible region.

## 1.2 Types of TCOs

TCOs are classified into two categories, i.e., *n*-type and *p*-type, depending upon the nature of conduction in the material. The detail explanation of both types of TCOs is discussed below:

### 1.2.1 *n*-type TCOs

High optical transparency in the visible region and high electrical conductivity at room temperature are usually believed to be incompatible within a single material. Since the material has a wide optical band equal to or more than 3.1 eV makes extrinsic doping very difficult. But TCOs can acquire high conductivity even close to that of metal ($10^4$ S/cm) and excellent transparency (80%) as that of ceramics in the visible region by introducing dopants and/or non-stoichiometry. This kind of technique was also employed in oxides of Cd, In, Sn, and Zn thin films by using different deposition techniques. Above mentioned materials are insulating in nature without extrinsic doping and acquired an optical band gap of ≥ 3 eV [16]. These materials were degenerately doped with appropriate countercation to lift the Fermi level into conduction band, demonstrating large carrier mobility and low absorption in the visible region of electromagnetic spectrum. An increase in carrier concentration will also increase the optical absorption, while improvement in mobility has no adverse effect on the transparency of materials. Therefore, the focus is to develop TCO material with large carrier mobility and high optical transparency. The very first *n*-type transparent conductor was realized in 1907 by Badekar in the form of CdO, whose thin film was deposited by sputtering of target Cd metal experienced incomplete thermal oxidation upon post-annealing heating in an ambient environment [7]. Hence, technological interest has been generated after the discovery of CdO as the first transparent conductor. Though CdO is not widely used nowadays as it is toxic in nature but still receives great interest due to its great carrier mobility. Afterward, extensive work has been done to develop better performance TCO so that they can render a wide variety of optoelectronic devices-based applications. There are various highly effective and commercially used TCOs available, such as Sn/F/Sb/Mo-doped $In_2O_3$, In/Al/F/B/Ga-doped ZnO, Sb/F-doped $SnO_2$, $MgIn_2O_4$, $Cd_2SnO_4$, Y-doped $CdSb_2O_6$, Sn-doped $CdIn_2O_4$, $Zn_2SnO_4$, $Zn_2In_2O_5$, $ZnSnO_3$, $ZnGa_2O_4$, $In_4Sn_3O_{12}$, Ge/Sn-doped $GaInO_3$, Sn-doped $AgInO_2$, $BaSnO_3$ etc. [4, 15–40]. In most of the *n*-type oxide, the ionic character of host acquires a valence band comprising O 2p orbital, and conduction band consist of metal ns orbital, which are spatially distributed, exhibit excellent *n*-type conductivity and large optical band gap as well. Out of these, ITO (Indium-tin oxide) is the best performance TCO reported in 1954, exhibits transparency of 80%, and conductivity of the order of $10^4$ S/cm [6]. Hence, its unique optoelectronic properties extend applications in solar cells and flat panel display devices as transparent electrodes [41]. Besides, ITO is considered as the best *n*-type TCO to date and accounts for more than 90% of the optoelectronic market due to its ease of fabrication along with high transparency and conductivity. However, due to low abundancy and regularly increasing price from the past three decades, forced researchers to explore other materials as possible alternatives to ITO.

### 1.2.2 *p*-type TCOs

The struggle for achieving *p*-type TCOs with comparable performance to *n*-type is due to their valence band structure. Because in *n*-type oxides, conduction band minimum (CBM) mainly consists of metal ns-orbitals. And it is a well-known fact that s orbital is highly delocalized or spatially distributed and may lead to enough hybridization even in the amorphous crystal structure. This highly dispersed or delocalized CBM results in the low effective mass of charge carriers (i.e., electrons) and great mobility. On the contrary, there is a lack of conductivity in *p*-type TCOs due to the following reasons: (i) production of charge carriers (i.e., holes) is limited by high formation energy of intrinsic acceptors that create holes like cation vacancies ($V_c$), oxygen interstitials ($V_i$) and (ii) low formation energy of intrinsic donors that can act as "hole killers" like oxygen vacancies ($V_o$) (iii) holes transport pathways in valence band maximum (VBM), as there is strong localization around oxygen 2p orbitals, which results in the larger effective mass of holes and impart low mobility (iv) Dispersion of valence band is so small that VBM level lies so deep, making hole doping unfavorable [9,42,43].

All the drawbacks mentioned above extend to the fact that the development of highly effective *p*-type TCOs is not a trivial task. But in 1997, Kawazoe et al. introduce the concept of chemical modulation of the valence band (CMVB) technique, as shown in figure 4 [42]. Following this concept, *p*-type TCOs conductivity can be improved by introducing appropriate countercation, whose energy-filled levels must be comparable or more than oxygen 2p levels. It can promote the delocalization around oxygen 2p orbitals by hybridizing metal-cation s, p or d orbital with oxygen 2p orbital to form a covalent bond and thus, lead to the dispersion of valence band minimum (VBM), which in turn, reduce the effective mass of holes and improve carrier's mobility.

The research for *p*-type TCOs was accelerated when a breakthrough was given by Kawazoe and co-workers in 1997, in which they invented a good performance *p*-type TCO in the form of $CuAlO_2$ [4]. In $CuAlO_2$ thin film, $Cu^+$ has $3d^{10}$ electronic configuration, and its energy level is just above the O 2p level, which is responsible for significant improvement in hole conductivity. Besides, the closed-shell configuration of $Cu^+$ ions avoided coloration and ascribed to a wide optical band gap. However, the electrical conductivity obtained from this material is still very low as compared to highly effective *n*-type TCOs. But this report makes other researchers realize that the development of better *p*-type TCOs is not an invincible goal. After that, a series of *p*-type TCOs were invented that can lead to the development of novel **"invisible electronics"** devices.

Following the concept of the CMVB technique, various oxide materials with degenerately doping were identified as *p*-type TCOs such as $CuMO_2$ (where M is Al, Mg, Cr, In, Sc, Y, and Ga) and $SrCu_2O_2$ [44–50]. The development of *p*-type TCOs demonstrated that the fabrication of transparent *p-n* junction is possible. After investigating the Cu-based *p*-type TCO materials, the concept of CMVB technique was also employed to chalcogens based materials. In this type of material, oxygen was intended to be replaced by chalcogens like S, Se, and Te. So that, there will be a hybridization between metal d orbital and chalcogen p-orbital. Layered $Sr_3Cu_2Sc_2O_5S_2$, LaCuOS, LaCuOSe and LaCuOTe were also identified as a potential candidate for *p*-type TCOs [51–55]. In addition, Mg-doped LaCuOS exhibits the highest *p*-type conductivity of 910 S/cm to date [56]. But its optical transparency is limited due to a low optical band gap of 2.8 eV. Lately, the CMVB technique was also employed on oxide materials with quasi-closed shell configuration like $d^6$ and $d^3$. This results in the realization of new TCO materials such as $ZnM_2O_4$ (where M is Co, Rh, and Ir) and chromium-based oxides [57,58].

**1.3 Why *p*-type TCOs are required?**

The history of TCO is more than a century ago, when Badekar invented the first TCO in the form of CdO in 1907 [7]. TCOs are extensively utilized in various optoelectronic applications. Still, active devices are absent such as bipolar diodes and transistors, due to lack of better performance *p*-type TCOs. Active electronic devices cannot be fabricated without *p-n* junctions. But the majority of TCOs available in the market are *n*-type in nature. Therefore, optoelectronic applications are limited to *n*-type TCOs based unipolar devices only. Furthermore, applications like solar water splitting and photovoltaics are also looking for *p*-type TCOs as electrodes for a better efficient collection of holes [59]. The difficulty encountered in *p*-type TCOs is intrinsic band structure of oxides. As oxygen is a relatively small atom with high electronegativity. Hence, it is responsible for the struggle to form a shallow acceptor level and results in a large effective mass of holes. However, attempts are also being made to develop acceptor level defects in the previously existing *n*-type TCOs for fabricating transparent *p-n* homojunction. But unfortunately, no TCO material has been invented yet, which can demonstrate ambipolarity.

Materials like ZnO and $SnO_2$ were tried to convert their conduction from *n*-type to *p*-type by introducing different dopants but failed. DFT study revealed that intrinsic defects like oxygen vacancies and cation interstitials act as donor defects, which abolish the *p*-type dopants [60–62]. Besides, Theoretical investigations have shown that band structure requirement for *n*-type and *p*-type dopability is completely contradictory to each other.

If *p*-type TCOs with equally good performance will be available, then transparent *p-n* junction and highly efficient energy circuits with more complexity can be fabricated. For instance, alike silicon technology, a transparent complementary metal-oxide-semiconductor (CMOS) device with effective performance and similar complex circuits can be realized. Hence, the development of highly effective *p*-type TCOs became a major research topic in **"invisible electronics"**.

### 1.4 Development of *p*-type TCOs

Since the groundbreaking work by Kawazoe et al. in 1997, there has been a surge in research on *p*-type transparent conducting oxides (TCOs) for applications in transparent electronics. This review investigated the scientific output in this field, focusing on the evolution of published papers between 1997 and 2024. Figure 5 illustrates a bar chart categorizing published research on *p*-type TCOs by material type: binary oxides, spinel oxides, chromium (Cr)-based oxides, and delafossite. Notably, the number of publications escalated significantly after 2003, exceeding 500 publications by 2010 and so on.

Initially, research focused immensely on delafossite structures; however, there has been a gradual shift towards exploring binary oxides over the years (discussed later). The increase in publications and the evolving focus of research show up the dynamic nature of the field and the ongoing efforts for highly effective *p*-type TCO materials.

This review provides a comprehensive survey of *p*-type TCO materials explored in previous studies. We have examined various host materials that have been investigated for their potential in *p*-type TCO applications. This examination considered crucial factors such as optical transparency, electrical conductivity, chemical stability, and the deposition processes of *p*-type TCO materials.

#### 1.4.1 Cu-based materials as *p*-type TCOs

$CuAlO_2$ material received great interest due to its wide optical band gap and significant *p*-type conductivity. After that, various Cu-based delafossite structures have been studied and their optoelectronic properties are highlighted in table 1. However, hole conductivity in $CuAlO_2$ was firstly reported in 1984 by Benko and Koffyberg [63]. But they have examined optoelectronic properties of powder sample compacted into a disk. The disk sample exhibited the resistivity of 5.9 Ω-cm with a very low value of Hall mobility of $10^{-5}$ $cm^2/Vs$ at room temperature. A detailed structural study of $CuAlO_2$ was reported by Ishiguro et al. [64]. They have also determined the electron-energy distribution in crystals of $CuAlO_2$ [65]. Thereafter, Kawazoe et al. in 1997 prepared a transparent thin film of $CuAlO_2$ material and studied its optoelectronic properties.

Following the CMVB technique by introducing $Cu^+$ cation, other Cu-based delafossite structures were reported as *p*-type TCO material in the form of $CuGaO_2$ and $CuInO_2$. The optical and electrical properties of $CuGaO_2$ thin film were studied [49]. As-deposited $CuGaO_2$ was found epitaxial in nature, confirmed by cross-sectional HRTEM. UV-Vis optical transmission spectra demonstrated that $CuGaO_2$ thin film was highly transparent in the visible region with an average transparency of 80%. The observed electrical conductivity at room temperature was very low compared to $CuAlO_2$. However, the conduction mechanism is similar in both materials, i.e., variable-range hopping.

Thereafter, the optoelectronic properties of doped $CuInO_2$ thin film was explored by introducing counter-cation and optimized thin-film deposition parameters [46]. Two different dopants, such as Ca and Sn were used, resulting in *p*-type and *n*-type thin film conduction. The PLD technique was used to deposit a thin film of Ca and Sn doped $CuInO_2$. All parameters were kept the same except $O_2$ partial pressure, which is 1 Pa for Ca $CuInO_2$ and 1.5 Pa for Sn doped $CuInO_2$ thin film, as the conductivity of Sn:$CuInO_2$ thin film is susceptible to $O_2$ partial pressure during deposition. Both Sn:$CuInO_2$ and Ca:$CuInO_2$ were highly transparent in the visible range with a direct optical band gap of 3.9 eV. However, thickness of Sn:$CuInO_2$ thin film was 280 nm and 170 nm for Ca:$CuInO_2$ thin film. Considering the electrical properties of *n*-type Sn:$CuInO_2$ and *p*-type Ca:$CuInO_2$ thin film, low conductivity value (of the order of $10^{-3}$ S/cm) was observed at room temperature. The nature of conduction was demonstrated by sign of Seebeck coefficient, which was -50 μV/K for Sn:$CuInO_2$ and +480 μV/K for Ca:$CuInO_2$ thin film. The substitution of $Ca^{2+}$ ions at $In^{3+}$ cationic sites result in *p*-type conductivity, whereas $Sn^{4+}$ ions at $In^{3+}$ attributed to *n*-type conduction.

They have also fabricated a transparent *p-n* homojunction composed of Sn:$CuInO_2$ and Ca:$CuInO_2$ films [66]. Initially, ITO film with a thickness of 600 nm was epitaxially grown on

(111) yttria-stabilized zirconia (YSZ) substrate by using the PLD technique. Subsequently, *p*-type and *n*-type $CuInO_2$ thin film having a thickness of 400 nm was deposited on ITO. At last, again ITO film (act as a cathode) deposited on multi-layered film with a thickness of 400 nm. The characteristics I-V curve of transparent *p-n* homojunction illustrated the rectifying behavior, i.e., current in forward biasing was 10 times higher than the reverse one in the voltage range from –4 to +4 V. The diode in thin-film form was also showed good transparency in the visible region, varying in the range of 60 to 80%. However, the transmission was very poor in the NIR region due to ITO carrier electrons.

As desired hole conductivity was not achieved in pristine Cu-based delafossite structure. Therefore, various dopants have been tried in Cu-based *p*-type TCOs such as Fe-doped $CuGaO_2$, Mg-doped $CuScO_2$, Mg-doped $CuCrO_2$, and Ca-doped $CuYO_2$ [67]. A thin film of $CuGa_{1-x}Fe_xO_2$ was deposited via PLD technique, and conductivity was found to be increased from 0.02 to 1 S/cm. The idea was to combine the optical properties of $CuGeO_2$ and electrical properties of $CuFeO_2$. As Benko et al. in 1986 already determined the optoelectronic properties of $CuFeO_2$ in powder form and found a minimum resistivity value of 0.11 Ω-cm at 2% of Mg-doped $CuFeO_2$, which was 0.65 Ω-cm in pure form with low mobility value of 0.27 $cm^2$/V-s. But on the contrary, a reduction in transparency was observed with Fe doping, and the optical band gap reduced from 4.3 eV to 3.4 eV, which is very near to the band gap of $CuFeO_2$ [68].

Transition metal (Sc) was also employed in Cu-based delafossite structure and prepared a thin film of $CuScO_{2+x}$ by sputtering technique [47]. Polycrystalline $CuScO_2$ film was found insulating in nature with a large optical band gap of 3.3 eV, illustrating higher optical transparency. Afterward, as-deposited thin films were post-annealed under the $O_2$ environment for 2 hours at 450 $^oC$. This oxygen intercalation results in an improvement in conductivity value to 15 S/cm (at room temperature). But transparency of thin film was degraded to 40% in the visible spectrum. However, they have also achieved a conductivity value of 30 S/cm by increasing intercalation pressure, as shown in figure 6. Still, this conductivity value is lower than *n*-type TCOs, but this report paved another way for increasing the *p*-type conductivity. Optoelectronic properties of oxygen intercalated Mg-doped $CuScO_{2+x}$ thin film was also examined [69]. Mg-doping improved the conductivity to 0.48 S/cm of $CuMg_ySc_{1-y}O_{2+x}$ pallet. On the other hand, oxygen intercalation enhanced thin film's *p*-type conductivity to 25 S/cm.

However, Mg incorporation didn't affect the optical transparency, but oxygen intercalation darkened the thin film.

After the transition metal, rare earth element (Y) was also employed by Jayaraj et al. in 2001 [48]. They have deposited the thin film of $CuY_{1-x}Ca_xO_2$ by thermal co-evaporation technique. They have prepared thin films on different substrates, i.e., glass, Si, and single crystal (100) MgO. As-deposited thin films were non-conductive, turn into conductive, followed by rapid thermal annealing in the $O_2$ atmosphere, and improved transparency as well. Still, no exciting results were obtained from this material, exhibited conductivity of 1 S/cm with Ca doping (which was 0.03 S/cm without Ca) and transparency varying between 40-50% in the visible range. Highly oriented thin film on the MgO substrate didn't illustrate any improved optoelectronic properties over polycrystalline thin films deposited on glass and Si substrate. They have also fabricated a transparent *p-n* diode in the form of glass/ITO/$Zn_{1-x}Al_xO$ /$CuY_{1-x}Ca_xO_2$/In by r.f. magnetron sputtering technique and demonstrated rectifying behavior.

In 2001, when a wide variety of Cu-based delafossite structures have been explored, Nagarajan et al. developed a p-type $CuCr_{1-x}Mg_xO_2$ material with the excellent value of hole conductivity [45]. They have synthesized $CuCr_{1-x}Mg_xO_2$ material, both in bulk and thin-film form. Undoped bulk $CuCrO_2$ material was found to be black, with an average value of conductivity. And conductivity observed in undoped $CuCrO_2$ thin film was of the order of 0.1 S/cm. When 5% Mg was introduced into $CuCrO_2$, its conductivity was increased by a factor of $10^{3,}$ i.e., 220 S/cm, which was highest at that time for any *p*-type TCO material. However, this material's optical transparency was low in the visible region, i.e., only 30% for 250 nm thick film. They have also post-annealed the thin film at different temperatures. No significant change in crystallinity, conductivity, and transparency has been observed upto 600 ºC. But at 900 ºC, transparency was improved to 40%, whereas the resistivity value also increased to ~1 Ω-cm. They have even attempted oxygen intercalation of as-deposited thin films, but that experiment turned out to be a failure. Due to the low lattice volume of $CuCrO_2$ compound, no oxygen intake was there at any temperature upto 1200 ºC, indicated by thermogravimetric analysis (TGA) results. After this temperature, oxygen intake leads to the formation of $CuCr_2O_4$.

In 2015, Farrell et al. prepared the Cu-deficient $CuCrO_2$ thin film by spray pyrolysis technique [70]. They have achieved maximum conductivity of 12 S/cm alongwith good transparency of 55% for 80 nm Cu-deficient $CuCrO_2$ thin film. The experiment of co-doping of Mg and N into $CuCrO_2$ thin films was conducted by Ahmadi et al. in 2018 [71]. Mg and N doped $CuCrO_2$ thin films were prepared by RF sputtering technique on the quartz substrate. A tremendous increase in electrical conductivity was observed from 0.01 S/cm (undoped $CuCrO_2$) to 277.7 S/cm (2.5% Mg, N doped $CuCrO_2$) alongwith better optical transparency of 69.1% in the visible region and band gap of 3.52 eV. Besides, 2.5% Mg, N co-doped $CuCrO_2$ thin film illustrated the highest hole conductivity to date despite having very poor Hall mobility of 0.006 $cm^2$/V-s.

Another Cu-based delafossite materials were also investigated by Nagarajan's research groups such as $CuM_{1-x}M'_xO_2$ (where M is Mg, Mn, Fe, Co, Ni, Zn and M' is Sb and V) [72]. Above mentioned Cu-based delafossite structures have been explored with doping (or even with co-doping) in the hope to obtain desired optoelectronic properties of *p*-type TCOs. But all materials were found to be highly resistive (in powder form) of the order of ~$10^6$ Ω-cm, and no optoelectronic studies on the thin films of materials have been reported so far. However, the same research group examined a thin film of Sn-doped $CuNi_{1-x}Sb_xO_2$ on a fused silica substrate. Though transparency in the visible region was found to be reasonable, the conductivity of thin film was 0.05 S/cm only at 10% of Sn doping.

Apart from the Cu-based delafossite structure, there is a material which is also Cu-based but exhibits a non-delafossite structure in the form of $SrCu_2O_2$. Kudo et al. have deposited a thin film of this material in 1998 [50]. Thin films of undoped $SrCu_2O_2$ and 3% K-doped $SrCu_2O_2$ were deposited. Both undoped and doped thin films (having a thickness of 150 and 200 nm, respectively) were highly transparent in the visible and infra-red range. The electrical conductivity of undoped thin film was found to be $3.9\times10^{-3}$ S/cm, increased to $4.8\times10^{-2}$ S/cm with 3% K-doping. The positive sign of Seebeck and Hall coefficients confirmed the *p*-type conductivity of both undoped and doped thin films.

In the recent past, various previously used Cu-based materials such as $CuCrO_{2+x}$, $CuCrO_2$, $CuGaO_2$ and $CuAlO_2$ were explored using different deposition techniques and parameters to surpass the previous results, as shown in table 1. However, the optical transparencies were found to be reasonably better than previous results, but no remarkable improvement in electrical conductivity was demonstrated [73–76].

### 1.4.2 Ag-based materials as *p*-type TCOs

As suggested by Kawazoe et al. to follow the concept of CMVB technique to enhance *p*-type conductivity by introducing suitable counter-cation with $nd^{10}$ configuration to promote delocalization in the valence band. After that, various Cu based *p*-type TCOs have been studied, as discussed above. But desired results were not achieved though they set a benchmark for other researchers that it is not an impossible task. Afterward, some Ag-based delafossite structures have been investigated as $Ag^+$ ions also have a $4d^{10}$ closed-shell configuration and its energy level is higher than $Cu^+$ ions that can promote more delocalization [77].

However, the synthesization of Ag-based delafossite structures is not straightforward through solid-state reactions and usually requires ion-exchange synthesis [67]. The first TCO thin film based on Ag-delafossite structure $AgInO_2$ was reported in 1998 by Otabe et al., but it was *n*-type [38]. They have deposited undoped and Sn-doped $AgInO_2$ thin films by r.f. sputtering technique. Both as-deposited thin films were post-annealed under $O_2$ flow at 500 ºC for 12 hours. Both thin films were highly transparent in the visible region, even upto near UV-region, and acquired a wide optical band of 4.2 eV for pristine and 4.4 eV for 5% Sn-doped thin film. Undoped $AgInO_2$ thin film exhibited very poor conductivity at room temperature, i.e., $10^{-5}$ S/cm. However, improvement in the Sn-doped $AgInO_2$ thin film was marvelous (6 S/cm). But still, it was very low as compared to the best-known *n*-type TCOs. A lower Hall mobility value of 0.47 $cm^2$/V-s was responsible for marginal *n*-type conductivity in Sn-doped $AgInO_2$ thin film. And *n*-type conduction was confirmed from the negative value of the Seebeck coefficient.

The synthesis and theoretical examination of $Ag_3VO_4$ was explored as potential *p*-type TCO candidate [78]. $Ag_3VO_4$ is nearly transparent according to theoretical calculations, with optical absorption at 2.6 eV, which is in close agreement with experimental results. But, with a conductivity limit of 0.002 S/cm and low hole concentration, making it undesirable for TCO applications.

In 2017, Renhuai Wei et al. reported the synthesis and characterization of *p*-type TCO thin films using $AgCrO_2$ (ACO) delafossite structure [79]. They found that the stoichiometric ACO thin films showed a preferred c-axis orientation, with optimal properties achieved at 500°C annealing temperature. To enhance conductivity, magnesium (Mg) doping was explored, resulting in *p*-type conductivity for $AgCr_{1-x}Mg_xO_2$ thin films, as depicted in table 2. As the Mg-content increased from 0.04 to 0.12, the conductivity of the thin films increases considerably from $3.1\times10^{-3}$ to $67.7\times10^{-3}$ S/cm. Beyond this, the conductivity somewhat decreases as the concentration of Mg increases. Hall measurements revealed changes in hole concentration and carrier mobility with varying Mg-content. Furthermore, the doped thin films exhibited high optical transmittance in the visible spectrum.

Lately, Keerthi et al. (2020) achieved excellent hole conductivity in $AgInGaO_2$ thin film [80]. They have deposited $AgGaO_2$ and $AgInGaO_2$ thin films evaporation technique followed by post-thermal annealing. The highest conductive value of 61 S/cm in Ag-deficient $AgInGaO_2$ thin-film, whereas optical transparency was only 25% despite having a wide optical band gap of 3.77 eV. Hall measurements results confirmed the *p*-type nature of as-deposited thin films, and at maximum conductivity value, Hall mobility and carrier concentration were 0.017 $cm^2$/V-s and $2.2 \times 10^{24}$ $cm^{-3}$, respectively.

### 1.4.3 Spinel-based oxides as *p*-type TCOs

The spinel structure has a general formula of $AB_2O_4$, where A is divalent cation occupies the tetrahedral site, and B is trivalent cation occupies the octahedral site. Windisch et al. (2001) deposited the first spinel oxide-based *p*-type TCO thin film by spin coating technique in the form of $NiCo_2O_4$ [81]. Pristine $NiCo_2O_4$ illustrated excellent transmittance in the visible range and found to be decreased when the ratio ($x$) of Co/(Co + Ni) was increased. In this material, the trivalent cation ($Co^{3+}$) substitution by divalent ion ($Ni^{2+}$) increased the hole conductivity, as shown in table 3. Minimum value of resistivity (~0.06 Ω-cm) was obtained at $x = 0.67$. Below this ratio, there is a formation of NiO, which increased the resistivity. Above that ratio, the compound was spinel only and increased Co-content also enlarged the resistivity value. However, several cations such as Al, Cu, and Zn were also introduced in the starting precursor solution. But obtained results were not upto the mark, and oxide films demonstrated a low conductivity value compared to pristine $NiCo_2O_4$ film.

Zinc-based spinel oxides have also been explored by different researchers in the form of $ZnM_2O_4$ (where M is transition-metal ions such as Co, Rh, and Ir). As these ions exhibit octahedral symmetry and acquire $nd^6$ electronic configuration, behave as "quasi-closed shell" configuration in a low-spin state. Mizoguchi et al. (2002) reported *p*-type conductivity in $ZnRh_2O_4$ thin film deposited using RFMS technique [82]. The thin film's conductivity was found to be 0.7 S/cm at room temperature with an optical band gap of 2.1 eV. Narushima et al. (2003) also deposited a thin film of $ZnRh_2O_4$ on a glass substrate by a similar technique, but it was amorphous. The optical band gap value was observed to be the same. But conductivity was increased to 2 S/cm at room temperature [83]. They have also fabricated a *p-n* junction in the form of Au/$ZnRh_2O_4$/$InGaZnO_4$/ITO. I-V characteristics demonstrated the rectifying behavior with an on/off current ratio of $10^3$ and threshold voltage was 2.1 V. In addition, Ohta et al. (2003) also fabricated a *p-n* junction in view of Au/$ZnRh_2O_4$/ZnO/ITO but in an epitaxial single-crystalline form [84]. The same rectifying behavior was observed with a threshold voltage of 2 eV. Conduction in amorphous $ZnRh_2O_4$ thin film was explained by Kamiya et al. (2005) [85]. They investigated that conduction in amorphous $ZnRh_2O_4$ type material emerges from the isotropic nature of spinel oxide's structure and networks of edge-sharing $RhO_6$ octahedra. And these are hardly effective by structural distortion and remain stable even in an amorphous state.

After this, H J Kim et al. (2004) deposited $ZnCo_2O_4$ thin film by reactive magnetron sputtering technique [86]. The optical band gap of the as-deposited thin film was 2.63 eV. Talking about the electrical properties, it was found to be entirely dependent on oxygen partial pressure. At low oxygen partial pressure, $ZnCo_2O_4$ thin film was *n*-type in nature with a carrier concentration of $1.37\times10^{20}$ $cm^{-3}$. But, as the $O_2$ partial pressure raised, conduction changed to *p*-type with a carrier concentration of $2.81\times10^{20}$ $cm^{-3}$ with Hall mobility of 0.2 $cm^2$/V-s, as shown in figure 7. Excess oxygen can result in diminishing O-vacancies and electron concentration. The activation energy was found to be 41 meV for *p*-type $ZnCo_2O_4$ thin film.

In 2007, Dekkers et al. deposited thin films of $ZnRh_2O_4$, $ZnCo_2O_4$, and $ZnIr_2O_4$ by PLD technique [57]. Among these three materials, $ZnIr_2O_4$ illustrated a wide optical band gap of 2.97 eV. However, the optical band gap of 2.74 eV was observed for $ZnRh_2O_4$, which was higher than the earlier reported value [82]. All thin films were *p*-type in nature, as determined by the Seebeck coefficient. The observed conductivities values for $ZnRh_2O_4$, $ZnCo_2O_4$, and $ZnIr_2O_4$ were 0.39/0.61, 2.75/2.83, and 3.39/2.09 S/cm, respectively (Previous conductivity value is for polycrystalline and later for the epitaxial thin film). After that, S Kim et al. (2010) deposited $ZnCo_2O_4$ thin film on the sapphire substrate by PLD technique [87]. They observed a high conductivity value of 21.8 S/cm (at room temperature) by optimizing oxygen pressure, which was higher than previously reported values [57]. The carrier concentration and Hall mobility values were found to vary with oxygen pressure. The optical band gap was found to be 2.3 eV. They have also deposited a *p-n* junction, but no satisfying results were found.

Using an inverse design approach, researchers identified $Cr_2MnO_4$ as a promising candidate for a new *p*-type transparent conducting oxide [88]. Using lithium as dopant, they significantly enhanced its *p*-type electrical conductivity at room temperature. High-temperature measurements confirmed its behavior as a band conductor, indicating potential improvements in its electrical properties with methods like single crystal growth or thin film deposition. However, the introduction of lithium reduced its optical transparency due to the formation of certain species and increased hole content within the material.

Hong-Ying Chen and Po-Chun Chen (2020) utilized the sol-gel process to fabricate thin films of *p*-type spinel $ZnCo_2O_4$ [89]. The films were produced through annealing at temperatures between 300 and 400°C in an oxygen atmosphere. The wide optical band gap of 3.95–3.99 eV was determined for the thin films. The films demonstrated resistivity of the order of $10^2$-$10^3$ Ω-cm. In the recent study, Z. Chi et al. (2021) successfully deposited both *p*-type $ZnGa_2O_4$ thin films by precisely controlling deposition parameters. The deposited *p*-type thin film was found to be highly resistive. But the hole conductivity in *p*-$ZnGa_2O_4$ was found to be enabled by $Zn_{Ga}$ anti-site defects. This study provides valuable insights into the conductivity mechanisms of $ZnGa_2O_4$, contributing to a deeper understanding of its electrical properties [90].

Various theoretical studies were also conducted on the electronic structure of spinel-based oxide materials [91–93]. But theoretical calculations were not well-consistent with experimental results, whether in terms of optical band variation or its magnitude. As discussed above, Dekkers et al. (2007) found that the optical band gap increased with a quantum number [57]. But different trends were observed in theoretical evaluation. In the case of spinel oxides, their *p*-type conductivity can be enhanced by creating acceptors like anti-site disorder and divalent cation vacancy. But theoretical studies explained that there was no noteworthy dispersion of VBM observed in spinel oxides, which lead to a large effective mass of holes and limiting hole conductivity. However, *p*-type conductivity was improved by oxygen intercalation. But this is significant upto some extent only, subsequently, it started diminishing.  Hence, spinel-based oxide materials do not show great potential as *p*-type TCOs.

### 1.4.4 Binary metal oxides as *p*-type TCOs

As discussed early in the introduction part, the first *p*-type TCO was NiO with a transparency and electrical conductivity of 40% and 7.1 S/cm (at room temperature), respectively. Table 4 illustrates the optical and electrical results of various binary oxide materials as *p*-type TCO. However, the optical band energy of single-crystal NiO was reported between 3.6 to 4 eV [5]. In NiO material, both factors, such as Ni vacancies and oxygen interstitial, are responsible for its *p*- type conductivity, as described by Antolini in 1992 [94]. Its valence band is comprised of hybridized Ni 3d orbital and O 2p orbital. Therefore, conduction is expected in NiO material. However, Li was also tried as a dopant to enhance the conductivity of NiO [95]. As substitution of $Li^{+}$ into $Ni^{2+}$ introduced a hole into the system, no satisfactory results were found. In 2014, Chen et al. deposited a thin film of NiO on the glass substrate [96]. They studied the effect of oxygen ion beam current on NiO thin films' optical and electrical properties. Stoichiometric NiO thin films were highly resistive of the order of $10^{13}$ Ω-cm (at 50% O-content). Reduction in resistivity was observed when excess oxygen was introduced during the growth process. The lowest resistivity value of 0.11 Ω-cm was determined at a discharge current of 0.42 A. Alongwith that, carrier concentration was also raised with oxygen ion beam discharge current, but Hall mobility decreased from 28.56 to 6.12 $cm^2$/V-s when the discharge current increased from 0.22 to 0.42 A. Considering the optical properties, NiO thin film exhibited average transparency of 69% in the visible range. But as the discharge current raised, transmittance was diminished to 22%.

In addition, Yang et al. (2012) also tried to enhance the hole conductivity in NiO through K-doping [97]. The highest conductivity of 4.25 S/cm was achieved at 25 at% of K-doping and noteworthy average optical transmission of 60% in the visible region. Hall measurements confirmed the *p*-type nature of as-deposited thin films. Another dopant Cu has also been introduced into *p*-type NiO thin films by Chen et al. in 2013 [98]. Cu-dopant was added at 10 and 18 at% into NiO thin films. The optical band gap energy was reduced with Cu-doping. The lowest value of resistivity was observed at 18 at.% of Cu-doping, i.e., 0.07 Ω-cm (see figure 8).

Despite having low conductivity and transparency, NiO is mostly used as a *p*-type TCO for transparent *p-n* junction and TFTs. One benefit of NiO is its simple rocksalt structure that may be found more compatible with other *n*-type TCOs. Ohta et al. (2003) deposited NiO/ZnO/ITO on (111) YSZ substrate by using the PLD technique [99]. The *p-n* diode illustrated the rectifying characteristics with a threshold voltage of 1 V and an ideality factor of 2. In 2009, Gupta et al. also fabricated the same *p-n* junction in the form of Au/ZnO/NiO/Au on via PLD technique [100]. They also showed rectifying behavior under the I-V characteristics curve with a barrier height of 0.33 eV and ideality factor of 4. The *p*-type NiO also finds applications in resistive-switching memory, electrochromic windows, and hole transport layer in thin-film solar cells [3,101]. Recently, begum et al. (2024) also tried to enhance the optoelectronic properties of NiO by doping with Zn and Mn, but no satisfactory results were obtained [102]. However, optical transmittance of NiO film was good as compared to previous results.

Moving ahead from NiO, some serious efforts have been made to develop ZnO as *a p*-type TCO. ZnO is *n*-type in nature and exhibits excellent optoelectronic properties. Hence, fabrication of transparent *p-n* homojunction can lead to a spectacular achievement in the field of **"invisible electronics"**. Doping elements such as B, F, Al, Ga, and In incorporated into ZnO lattice lead to *n*-type conductivity, and they are widely used in optoelectronic applications [16,103]. Lander performed the very first try in 1960 [104]. He doped Li into ZnO and studied its reactions as donor and acceptor. The main conclusion drawn from this report was that too high oxygen pressure is needed to develop *p*-type conductance in ZnO, which is impractical.

Yuichi Sato and Susumu Sato made another considerable effort in 1996 by mixing oxygen and nitrogen gas during ZnO thin film deposition on a sapphire substrate using reactive evaporation technique [105]. Films deposited under the $O_2$ environment were highly c-axis oriented and exhibited greater transparency in the visible region. Whereas the film deposited in the $N_2$ environment was amorphous in nature and reddish-brown in color with very poor transmittance. After that, mixing of both gases were done at a ratio of $N_2/(N_2 + O_2)$ 25, 50, and 75% at two different deposition temperatures of 300 and 400 $^{o}$C. But no noticeable change in transmittance spectra has been observed. Hall mobility was increased, but all thin films were *n*-type in nature.

Following the above concept, Minegishi et al. demonstrated the realization of *p*-type ZnO film on c-cut sapphire substrate by using a chemical vapor deposition technique in 1997 [106]. They illustrated the simultaneous incorporation of ammonia ($NH_3$) in carrier hydrogen and the surplus of Zn in ZnO powder. However, the resistivity was found in the order of ~100 Ω-cm. They suggested a model that introducing nitrogen into ZnO can lead to *p*-type with reasonable conductivity by optimizing temperature for thermal annealing. Similar electrical properties of undoped ZnO were observed with N-doing without excess of Zn. Also, they did not detect the presence of nitrogen in ZnO thin film at Zn/ZnO ratio of 0 mol%, indicating that incorporating N into ZnO lattice is not easy. However, at Zn/ZnO ratio of 10 mol%, the existence of N was confirmed. Hence, excess Zn was required to add nitrogen, as Zn likely to be combined with oxygen rather than nitrogen. sThe carrier density was found to be decreased and conduction transformed into *p*-type. This indicated the successful substitution of oxygen (O) by nitrogen (N) and results in the formation of acceptor level, estimated at nearly 0.1 eV above the top of the valence band. The activation of acceptor depends upon the substrate temperature region as low-temperature results in electrically inactive acceptor and intermediate temperature lead to activation of acceptor and reduced carrier density, which in turn, convert conduction into *p*-type. But high temperature broke the N-H bond of ZnNH before the addition of nitrogen into ZnO film. The control of temperature in that narrow range (where conduction nature is inversed) is complicated. That's why the resistivity of N-doped ZnO films was found to vary with substrate temperature. High resistive samples were obtained between 650 to 750 ºC. Such high resistivity value is not significant for optoelectronic applications.

Guo et al. (2000) also introduced nitrogen into ZnO thin film deposited using the PLD technique [107]. During the deposition, pure Zn metal was ablated under $N_2O$ plasma environment. They were successful in achieving the *p*-type conduction in ZnO thin films by optimizing $N_2O$ partial pressure and micro-wave input power. A significant value of resistivity was achieved. The conduction nature of *p*-type ZnO was confirmed by the Seebeck coefficient. Optical transparency achieved in *p*-type ZnO was very similar to *n*-type ZnO.

Another homoepitaxial N-doped ZnO thin film was deposited by Look et al. (2002) using the molecular beam epitaxy [108]. Both $O_2$ and $N_2$ gases were introduced (with varying ratios) in r.f. plasma source. A thin film of N-doped ZnO was deposited on Li-doped, semi-insulating ZnO substrate to prevent the influence of conduction from the substrate. The resistivity value of 40 Ω-cm along with carrier concentration and Hall mobility of $9\times10^{16}$ $cm^{-3}$ and 2 $cm^2$/V-s was obtained, respectively. No optical properties of as-deposited thin films were studied in this report.

Ammonia doped ZnO effort was also made by Wang et al. (2003). They deposited thin films of $NH_3$-doped ZnO by metalorganic chemical vapor deposition on $<02\bar{2}4>$ oriented sapphire substrate [109]. They also demonstrated hole conductivity in ZnO thin film with a resistivity of 102 Ω-cm and carrier concentration and Hall mobility of $1.69\times10^{16}$ $cm^{-3}$ 3.6 $cm^2$/V-s, respectively. The mechanism behind ZnO film's conductivity is that the concentration of N-acceptor was more than self-compensated native donors such as Zn-interstitials and O-vacancies. But after a certain extent, conduction nature again reverted to *n*-type with increasing $NH_3$ flux. Because hydrogen content was also rising with $NH_3$ flux, and hydrogen atom can also act as a donor. No comments on the optical properties of $NH_3$-doped ZnO films were highlighted in this report.

Moving ahead of N-doping, Ryu et al. (2000) tried arsenic (As) as a dopant to introduced hole conductivity in ZnO [110]. Thin films of As-doped ZnO were grown on (001) GaAs substrates under $O_2$ environment by PLD technique. Dopant arsenic (As) act as a better acceptor, and its binding energy was found to be near about 100 meV. Post-annealing of thin films demonstrated the conversion of *p*-type conductivity from *n*-type due to diffusion of As-atom from the substrate to ZnO thin film. The *p*-type ZnO films exhibited carrier density around $10^{18}$-$10^{21}$ $cm^{-3}$, and Hall mobility was in the range of 0.1-50 $cm^2$/V-s. However, these values have significant uncertainties, as the contribution was not well-understood from interface layer between GaAs substrate and ZnO thin film. They have also not reordered optical properties of as-deposited and post-annealed thin films.

The concept of co-doping was also tried to develop ZnO as a *p*-type TCO. Yamamoto and Yoshida (1999) performed a first-principle study on Ga and N doped ZnO, in which Ga was assumed to act as donor and N as acceptor [111]. However, the concentration of N was twice that of Ga to achieve *p*-type conductivity in ZnO. The idea behind this approach was to decrease the repulsive N-N interaction, which in turn, creates a shallow acceptor level. Various reports were published by researchers following the above approach and claimed the lack of reproducibility of *p*-type ZnO using the co-doping technique [112–114]. The conclusion drawn from the above reports is that production of *p*-type ZnO is not a trivial task. Even hole conductivity obtained from the above reports was not outstanding from the application point of view. Furthermore, the main concern about its reproducibility may be due to lack of using proper growth conditions, or there may be a formation of unwanted impurity in ZnO matrix.

After this, a new *p*-type TCO material came into the picture in the form of SnO due to its abundancy, non-toxic nature, and comparatively high hole mobility. Various theoretical studies on the band structure of this material were investigated [115–117]. They revealed greater hole mobility in SnO material, due to its valence band structure and low-defect formation energy of Sn vacancies ($V_{Sn}$). The valence band of SnO mainly consists of hybridization of Sn 5s and O 2p orbitals, which lead to the large dispersion of VBM and Sn vacancies act as acceptors. All these intrinsic properties motivated other researchers to explore its optoelectronic properties by depositing a thin film of SnO. However, before investigating the *p*-type SnO, most of the research activities were attentive to $SnO_2$ material, which is *n*-type in nature and acquires applications like active electronics, transparent passive, and gas sensing [118–120]. For the very first time, Ogo et al. (2008) deposited epitaxial thin films of SnO and also fabricated a better top-gated *p*-type TFTs based on SnO [121]. The as-deposited thin film illustrated an optical band of 2.7 eV. They achieved Hall mobility of 2.4 $cm^2$/V-s at room temperature. Epitaxial SnO channels in TFTs demonstrated a reasonable performance with on/off current ratio of $10^2$, a threshold voltage of 4.8 V, and field-effect mobility of 1.3 $cm^2$/V-s.

Liang et al. (2010) also deposited a thin film of SnO on quartz and silica substrate followed by rapid-thermal annealing via e-beam evaporation technique [122]. They have also fabricated a *p*-type TFT using SnO as a channel. But not satisfactory results were obtained. In the same year, Fortunato et al. (2010) deposited thin films of $SnO_x$ (where $x$ <2) and bottom gated TFTs using RFMS technique [123]. XRD patterns determined the polycrystalline thin films, ascribed to β-Sn and α-$SnO_x$ phases. Average transparency was around 85% in the visible and IR range. Electrical resistivity varied between the range of 10-100 Ω-cm. $SnO_x$ based TFT demonstrated on/off current ratio of $10^3$ with field-effect mobility of >1 $cm^2$/V-s.

Besides, Kwok et al. (2022) investigated Ga and Na doping to improve conductivity in SnO thin films [124]. While Ga doping increased conductivity, Na doping achieved significantly lower resistivity but with lower transparency (<50%). Both dopants act as effective acceptors. The study suggested that Na-doped SnO has potential for TCO applications, but further research is needed to improve transparency.

SnO material can be a suitable candidate for *p*-type TCO to acquired transparent devices-based applications. But the main concern is about its stability, as it can be simply oxidized to $SnO_2$ (*n*-type) or reduced to metallic Sn. Therefore, it is challenging to grow single-phase SnO thin film from the device fabrication point-of-view.

Although $SnO_2$ is *n*-type in nature, efforts were made by various researchers to convert its conduction into *p*-type by introducing Al, Ga, and In cations at the Sn-lattice sites, as shown in figure 9 [62]. Varley et al. (2009) performed the first-principle study and revealed that $SnO_2$ material has low formation enthalpy, i.e., -5.42 eV (lower than ZnO). Besides, $SnO_2$ also has smaller ionization potential as compared to ZnO [125]. After that, Singh et al. (2008) concluded that $SnO_2$ material acquires significant chances to act as *p*-type TCO to render optoelectronic applications [126]. However, before these theoretical studies, Ji et al. in 2003 already reported the *p*-type conductivity in In-doped $SnO_2$ thin films deposited on quartz substrate using dip-coating technique [127]. All thin films were found to be *n*-type below 450 $^oC$ processing temperature, and above this, conduction changed to *p*-type. The lowest value of resistivity (i.e., 20.4 Ω-cm) was determined at 525 $^oC$ processing temperature with carrier concentration and Hall mobility of $1.85\times10^{18}$ $cm^{-3}$ and 1.57 $cm^2/V$-s, respectively. All thin films demonstrated the large optical band gap of 3.8 eV. Besides that, Lekshmy and Joy (2014) also shown optoelectronic properties of In-doped $SnO_2$ thin films by the same technique on same substrate as that of Ji et al. (2003) [128]. Upto 3 mol% In-doping into $SnO_2$, films exhibited *n*-type conductivity, whereas above 5 mol%, *p*-type conducting behavior was observed. Average transparency of more than 80% was reflected in UV-Vis spectra.

Huang et al. (2007) also reported the *p*-type conductivity in $SnO_2$ thin films with Ga-doping [129]. Considering the optical properties, thin film was highly transparent in the visible region with an average transmittance of >85%. On the other hand, the nature of conduction depended upon the oxidation temperature; too low or high temperature was not favorable for hole conductivity. They have achieved the *p*-type conductivity in the optimized temperature range of 600-650 $^oC$.

The rare-earth element (Eu) was also doped into $SnO_2$ thin film [130]. Undoped $SnO_2$ thin film was *n*-type in nature, but the inversion of conduction was observed with Eu-doping. The minimum value of resistivity (15.9 Ω-cm) was observed at 0.2 mM of Eu-doping. Carrier concentration was increased as $Eu^{3+}$ cations at $Sr^{4+}$ lattice sites act as acceptors in $SnO_2$ thin films. Simultaneously, a drastic reduction in Hall mobility was demonstrated from 57.2 to 2.71 $cm^2/V$-s due to the degradation of crystallinity and charge carrier's scattering, as depicted in figure 10. Talking about the optical properties, increment in the transparency was illustrated with Eu-doping, due to the Moss-Burstein shift. They have also prepared a bottom-gated TFT by using Eu-doped $SnO_2$ thin film as a *p*-type channel. The I-V characteristics curve demonstrated an ohmic behavior due to the low resistivity of thin film.

Apart from the above dopants, Al-dopant has also been tried to achieve better *p*-type conductivity in $SnO_2$ material. Mohagheghi and Saremi (2004), Ahmed et al. (2006), and Ravichandran and Thirumurugan (2014) demonstrated *p*-type conductivity in Al-doped $SnO_2$ thin films deposited on glass substrates [131–133]. Mohagheghi and Saremi (2004) and Ravichandran and Thirumurugan (2014) used the spray pyrolysis technique, whereas Ahmed et al. (2006) used the dip-coating technique for the growth of Al-doped $SnO_2$ thin films. They have observed that undoped and Al-doped $SnO_2$ thin films (at low Al-content) exhibited *n*-type conductivity. After a particular doping-content of Al, conduction was changed to *p*-type. Among all three reports, Mohagheghi and Saremi (2004) illustrated the lowest resistivity of 0.036 Ω-cm at 8.4 at% of Al-doping with carrier concentration and Hall mobility of $6.7\times10^{18}$ $cm^{-3}$ and 25.9 $cm^2$/V-s, respectively. However, Ravichandran and Thirumurugan (2014) observed the better carrier concentration ($3\times10^{19}$ $cm^{-3}$) and Hall mobility (more than 50 $cm^2$/V-s) value, but still, the resistivity was found to be high (0.38 Ω-cm). It is worth mentioning that Mohagheghi and Saremi (2004) and Ravichandran and Thirumurugan (2014) showed a decrement in the optical band gap with Al-doping. But, in Ahmed et al. (2006), optical band gap was increased from 3.87 to 4.21 eV. They have claimed that due to a decrease in particle size and some traces of Al-O bonding (particularly at grain boundary) observed in FTIR spectra responsible for the widening of optical band gap. The concept of co-doping was introduced into $SnO_2$ material to improve its optoelectronic properties. Duong et a. (2020) deposited Al and N doped $SnO_2$ thin films DC magnetron sputtering technique [134]. The optimized content of 6 wt% $Al_2O_3$ (with a mixture of 50% $N_2$ and 50% Ar gas) demonstrated the minimum resistivity value of $7.1\times10^{-3}$ Ω-cm with high Hall mobility and carrier concentration of 14.1 $cm^2$/V-s and $6.24\times10^{19}$ $cm^{-3}$, respectively. Apart from that, good optical transparency ~80% observed in as-deposited thin films.

Another binary oxide material explored as *p*-type TCO was $Cu_2O$. In 1966, Fortin et al. demonstrated an excellent value of Hall mobility in $Cu_2O$, i.e., 100 $cm^2$/V-s [135]. Because its VBM mainly consists of Cu 3d and O 2p orbital, which lead to the delocalization of holes and low holes' effective mass. Furthermore, Cu cations have tetrahedral coordination with O-ligands, which in turn, decreases the nonbonding nature of $O^{2-}$ anions and promotes further delocalization. Low formation energy of Cu-vacancies is responsible for its *p*-type nature [136–139]. Matsuzaki et al. (2008) reported the growth of *p*-type $Cu_2O$ thin films as TFTs [140]. The films were grown by PLD technique and reflected high Hall mobility of 90 $cm^2$/V-s at room temperature. But results obtained from TFTs were not reasonable with low on/off current ratio and field-effect mobility. Li et al. (2009) also prepared thin films of $Cu_2O$ by using RFMS technique [141]. They have investigated the effect of low-temperature buffer (LTB) layer of $Cu_2O$ on the deposition of $Cu_2O$ thin films and observed increment in grain size. They have achieved Hall mobility of 256 $cm^2$/V-s by optimizing the flow rate of oxygen during deposition (illustrated in figure 11), which is the highest reported value to date. No concern about optical properties was highlighted in this report.

Various deposition techniques have been employed to study the electrical properties of $Cu_2O$. There is only material (i.e., $Cu_2O$) whose Hall mobility varied from 0.05 to 256 $cm^2$/V-s. Different deposition techniques, substrates, and growth parameters are responsible for this massive fluctuation [58,141,142]. But very few reports on its optical properties are available in the literature. This drawback is the primary concern. Regardless of having an excellent mobility value, this material cannot be found suitable for **"invisible electronics"** due to its low optical band gap of 2.4 eV. This small optical band gap of $Cu_2O$ is due to the direct interaction of neighboring Cu $3d^{10}$ electrons. As discussed in $CuAlO_2$, the presence of $Al^{3+}$ ions reduced the $Cu^+$-$Cu^+$ interaction along c-axis, which is responsible for its large band gap. But this is not the case with $Cu_2O$, and not so much effort has been made to enlarge the band gap of $Cu_2O$.

### 1.4.6 Chalcogen material as *p*-type TCOs

Chalcogenides such as S and Se were employed in layered structure of (RO)CuCh (where R is rare-earth elements and Ch is chalcogenides) thin films to develop *p*-type TCOs. The CMVB technique employed for developing effective *p*-type TCOs also extended to these kinds of materials. As the energy-filled level of S 3p and Se 4p lying below than O 2p level, responsible for the widening of optical band gap. Furthermore, due to the presence of rare-earth element (La), which is highly electropositive, there will be a high energy shifting of conduction band minimum (CBM). The number of Cu-S bonds in unit cells also decreases, thereby promoting dispersion at the top of the valence band maximum (VBM) and improving mobility, clearly depicted in table 5. Therefore, this type of material showed great potential in the field of **“transparent electronics”**.

This type of material was firstly developed by Palazzi in 1981, and its hole conductivity was reported after a decade by Ishikawa et al. (1991) and Takano et al. (1995) [143–145]. Ueda and co-workers were the first ones who prepared a thin of thin LaCuOS [53]. Thin films deposited was highly transparent in the visible region (>70%). They have also doped Sr into LaCuOS and conductivity was found to be increased from $1.2\times10^{-2}$ to 0.26 S/cm. Similar Sr-doped LaCuOS thin films were grown by Hiramatsu et al. in 2002 [146]. As-deposited thin films were post-annealed in an evacuated silica tube comprising a small quantity of LaCuOS powder.

The electrical conductivity of films was increased from $6.4\times10^{-5}$ S/cm (undoped) to 20 S/cm (3% Sr doped). Optical transmittance of > 60% was reflected in the visible and NIR region.

The same group made another effort in 2003 by substituting Se at the S anionic site in LaCuOS thin film [54]. Hall mobility was found to be increased with Se doping and reached a maximum value of 8 $cm^2$/V-s. Epitaxial LaCuOSe thin film exhibited the maximum conductivity value of 24 S/cm. As a significant change in carrier density was observed, improved mobility was responsible for high conductivity value. For the sake of increasing the conductivity value of LaCuOSe, various divalent cations were also tried (such as $Mg^{2+}$, $Ca^{2+}$, and $Sr^{2+}$). Mg-cation came out as an effective dopant and increased the conductivity of epitaxial 20% Mg-doped LaCuOS thin film to 140 S/cm, due to increased carrier concentration. On the other hand, mobility was decreased to 4 $cm^2$/V-s. The determination of the optical properties of as-deposited thin films was missing in this report [56].

In 2020, Zhang et al. substituted rare-earth element (Y) at the rare-earth (La) cationic site in LaCuOS material [147]. They have doped Y into LaCuOS at various contents. Electrical conductivity was found to have increased from 5.6 S/cm ($x$ = 0) to 89.3 S/cm ($x$ = 0.25), which is highest till date in case of LaCuOS material. Furthermore, Hall mobility and carrier concentration were improved with Y-doping. On the other hand, optical transparency was slightly decreased with Y-doping, i.e., from 81.2% to 76.1%. They have also fabricated diode in the form of Y-doped LaCuOS/Al-doped ZnO and results in very high rectifying having on-to-off ratio of 300 at ±3V and threshold voltage around 1.1 V alongwith very low leakage current of $5\times10^{-7}$ A. Apart from that, this *p-n* junction exhibited greater transparency of more than 65% in the visible range. Hence, this material can be a potential candidate in next-generation **"transparent electronics"**.

Apart from La, various rare-earth elements such as Y, Pr, Nd, and Gd have also been explored [148]. But they have studied structural properties and band structure. Their optoelectronic properties still need to be investigated.

### 1.4.7 Cr-based oxides as *p*-type TCOs

Recently, chromium oxide ($Cr_2O_3$) received tremendous attention as *p*-type TCOs. Due to low formation energy of Cr-vacancies ($V_{Cr}$), $Cr_2O_3$ is *p*-type in nature [149]. The optical band gap energy of $Cr_2O_3$ is 3.1 eV. Apart from that, two kinks were always observed at 2 and 2.6 eV, which ascribed to dipole-forbidden d-d transition [150]. However, these transitions are very weak in thin-film form and barely affect the transparency in the visible region. Stoichiometric $Cr_2O_3$ is insulating in nature, and its conductivity was found to be increased by introducing suitable dopants. In 1996, Uekawa and Kaneko investigated that Li, Mg, and Ni dopants can increase conductivity [151]. They have illustrated only resistance-temperature study and observed that resistance decreases with Li, Mg, and Ni doping. The reduction of resistance in Li-doped samples was marginal. On the other hand, a rise in absorption intensity was demonstrated with Mg-doping, as addition of Mg into $Cr_2O_3$ leads to the formation of mix valence state of $Cr^{3+}$ and $Cr^{6+}$ ions. And the charge was also transferred from $O^{2-}$ 2p to $Cr^{6+}$ 3d, known as ligand-to-metal charge transfer (LMCT), and the color of as-deposited thin films changed from green to brownish.

Arca et al. (2011) were the first ones who investigated the optoelectronic properties of Mg-doped $Cr_2O_3$ thin films [152]. They have also tried to prevent the formation of Cr mixed-valence state by introducing co-doping of nitrogen alongwith Mg. The optical transparency of undoped and Mg-doped $Cr_2O_3$ thin films was very ordinary, as shown in table 6. However, increment in transparency was demonstrated with Mg and N co-doping, and maximum transmittance of 65% was observed in the NIR region, as shown in figure 12. On the other hand, electrical resistivity was drastically reduced from 400 Ω-cm to 15 Ω-cm. They have achieved the best results with an optimized ratio of Cr:Mg 9:1 and Cr:N 1:4. The lowest resistivity achieved in this report was 3 Ω-cm. All thin films were *p*-type in nature.

This study was further extended by the same group in 2013. They have investigated the role of different chemical precursors on the optoelectronic properties of Mg and N-doped $Cr_2O_3$ thin films [150]. The smallest value of resistivity 4 Ω-cm was determined using nitrate precursor at 0 pH value, whereas thin films deposited using chloride and acetate precursor results in high resistive film. The optical band gap was increased with N-doping because introduction of nitrogen reduced the concentration of $Cr^{6+}$ ions. They have also explained the mechanism of conduction in epitaxial Mg-doped $Cr_2O_3$ thin films deposited using molecular beam epitaxy (MBE) technique followed by post-annealing in an oxygen environment [153]. They have reported that conductivity of as-deposited thin films can be increased with post-thermal treatment under oxygen flow, illustrated in figure 13. However, conductivity of epitaxial Mg-doped $Cr_2O_3$ thin films wouldn't surpass their previous result.

A similar group in 2017 reported the optical and electrical properties of Ni-doped $Cr_2O_3$ thin films deposited using the PLD technique [154]. Significant *p*-type conductivity in $Cr_2O_3$ thin film was achieved by Ni doping (28 S/cm). They have also performed DFT calculations, which stated that the introduction of Ni promoted the delocalization of VB of $Cr_2O_3$, and it is experimentally proved that Ni has a higher solubility as compared to Mg. Due to these two advantages, Ni was found to be more effective dopant in $Cr_2O_3$ material. However, optical transparency was not too high, varying between 35 to 55% in the visible region.

We have also used $Cr_2O_3$ as a host material and Mg, Al, and Ni as substituents to enhance the optoelectronic properties of $Cr_2O_3$. We have deposited thin films of $Cr_2O_3$, $Mg_xCr_{2-x}O_3$ ($x$ = 0.1, 0.2, and 0.3), $Al_xCr_{2-x}O_3$ ($x$ = 0.1, 0.2, and 0.3) and $Ni_xCr_{2-x}O_3$ ($x$ = 0.03, 0.05, and 0.1) nanomaterial using PLD technique [155,156]. All thin films were found single crystalline and epitaxially grown on c-cut sapphire substrates. And crystal structure of $Cr_2O_3$ thin films remains unaffected with Mg, Al, and Ni-substitution. All thin films illustrated the excellent optical transparency in the visible range, suitable for optoelectronic applications. The incorporation of Mg and Ni leads to the improvement in the electrical conductivity of $Cr_2O_3$ thin film, without much degrading its optical transparency. On the contrary, Al-substitution into $Cr_2O_3$ thin film further degraded its electrical properties. Comparing the $Ni_xCr_{2-x}O_3$ and $Mg_xCr_{2-x}O_3$ thin films, nickel came out as an effective dopant for improving the electrical conductivity of $Cr_2O_3$ material. Because its higher energy-filled 3d orbital can destroy the robust localization around oxygen 2p states, reduces the effective mass of holes and increases conductivity. It was also reflected in Hall measurement results, where hole mobility was increased to 2 $cm^2$/V-s from 0.1 $cm^2$/V-s (pristine $Cr_2O_3$ thin film). However, carrier mobility was also found to be improved to 0.4 $cm^2$/V-s in $Mg_xCr_{2-x}O_3$ thin films due to the hybridization of Mg and O outer shell orbitals. Besides, Hall measurement results confirmed the *p*-type conductivity in all thin films.

Apart from $Cr_2O_3$ material, perovskite structured material such as $LaCrO_3$ has also been investigated as a *p*-type TCO. Because excellent thermal and chemical stability of perovskite structure makes the hole doping easy and can be fabricated with other *n*-type perovskites TCOs to form transparent *p*-*n* junctions. And that can be an ideal candidate for a new class of solar cell materials, i.e., perovskite solar cells. However, $LaCrO_3$ is insulating in nature with an optical band of 4.6 eV. Apart from optical transition, two more inter/intra-band transition occurred in $LaCrO_3$ material, first from intra Cr ($t_{2g}$) to ($e_g$) at 2.7, 3.6 eV and other from inter Cr ($t_{2g}$) to ($t_{2g}$) at 4.4 eV [157].

In 2015, Zhang et al. demonstrated the new *p*-type TCO by doping Sr into $LaCrO_3$ thin film [158]. Undoped thin film results as an insulating sample. Hole conductivity was increased with Sr-doping content, and maximum conductivity of 54 S/cm was achieved at $x = 0.5$ of Sr-doping. Both carrier concentration and Hall mobility value improved with Sr-content. On the other hand, optical transparency was reduced from 69.1% (undoped) to 42.3% ($x = 0.5$). In addition, they have also deposited a thin film of Sr-doped $LaCrO_3$ at $x = 1$ of Sr-content. Hole conductivity was increased exceptionally to 720 S/cm, while optical transmittance was further diminished to 29%.

Recently, Machado et al. (2023) presented a significant advancement in the fabrication of LSCO ($La_{0.75}Sr_{0.25}CrO_3$) thin films through a cost-efficient chemical route [159]. The authors demonstrated that careful manipulation of solution chemistry leads to dense, epitaxial films with a slight strain, achieving approximately 67% optical transparency in the visible spectrum, comparable to leading transparent conducting oxides (TCOs). However, the films exhibit higher electrical resistivity attributed to structural defects, despite chemical doping and strain manipulation. To enhance the film performance, the authors proposed various strategies, including growth on lattice-matched substrates, minimizing antiphase boundaries, and optimizing the La/Sr ratio to boost conductivity without compromising optical transparency. Overall, this research contributed significantly to the understanding and development of LSCO as a *p*-type transparent conducting perovskite oxide, highlighting the potential of solution processing as a cost-effective method for fabricating functional complex oxides.

Considering this Sr-doped $LaCrO_3$ as new *p*-type TCO material, dabaghmanesh et al. performed the first principle study to demonstrate its electronic structure and formation energy of different point defects in *p*-type $LaCrO_3$ material [160]. They have computed that deposition of Sr-doped $LaCrO_3$ thin film in an oxygen-rich environment leads to the creation of shallow acceptors with smallest formation energy and results in better *p*-type conductivity. Whereas in poor oxygen environment, shallow acceptors were compensated by various intrinsic defects such as O-vacancies and Cr on the O-site. They have also determined that divalent cations like Ca and Ba can also be an effective dopant for improving the electrical properties of $LaCrO_3$, as did by Sr-doping. Both Ca, and Ba dopant has already been tried earlier by Ong et al. (2007) and Jiang et al. (2008) for solid oxide fuel cell applications [161,162]. They demonstrated that Ca-doped $LaCrO_3$ samples result in better electrical conductivity as compared to Ba-doping. It is worth mentioning that they have studied the electrical conductivity of Ca and Ba-doped $LaCrO_3$ samples in powder form only. Still, there are no experimental results based on the optoelectronic properties of Ca and Ba-doped $LaCrO_3$ thin films reported yet.

### 1.4.8 Some other oxide materials as *p*-type TCOs

Some oxides other than the above-mentioned class of materials have been explored to develop highly effective *p*-type TCOs. Aksit et al. (2014) deposited thin films of $Ca_3Co_4O_9$ using spin-coating technique [163]. Reasonably good transparency of 67% was observed in the visible range while in the infra-red region, it was about 85%. The electrical conductivity of 18 S/cm was demonstrated at room temperature. However, the observed hole conductivity was much lower than the earlier reported value, which may be due to the nanoporous thin film [164]. Previously, Shikano and Funahashi (2003) demonstrated the excellent conductivity value of ~ 500 S/cm (at room temperature) for in-plane single-crystalline $Ca_3Co_4O_9$ thin film, deposited via modified strontium chloride flux technique [165]. No study was performed on the optical properties of as-deposited thin film. Another Co-based *p*-type TCO in the form of $Bi_2Sr_2Co_2O_y$ was grown epitaxially on $LaAlO_3$ substrate by using the PLD technique. Wang et al. (2014) demonstrated the excellent value of resistivity of 5.5 mΩ-cm with reasonable optical transparency of 51% in the visible range [166].

A mixture of two oxides, i.e., $In_2O_3$ and $Ag_2O$ thin films, was deposited using RFMS technique [167]. The electrical resistivity value was low as 0.0088 Ω-cm at 50% of $Ag_2O$ content, with high Hall mobility and carrier concentration of 17 $cm^2$/V-s and $4.2\times10^{19}$ $cm^{-3}$, respectively. Regardless of having excellent electrical properties, this material cannot be suitable for optoelectronic applications because of its poor optical transparency. Also, Chen et al. (2011) introduced In-doped $MoO_3$ *p*-type TCO in both single crystal nanowire and amorphous film forms [168]. The study demonstrated impressive optical transparency and low resistivity (as low as $5.98\times10^{-4}$ Ω cm) for 80 nm thin films across a broad wavelength range (400-800 nm), making them highly suitable for photovoltaic applications. Notably, amorphous films maintain excellent electrical properties even on flexible polyimide substrates, showcasing their potential for flexible electronics. Furthermore, the fabrication of *p*-$MoO_3$:In/i-ZnO/*n*-AZO devices emphasizes their applicability in all-transparent flexible electronic applications.

Another exciting material as highly effective *p*-type TCO was introduced by Hu et al. in 2019 [169]. They have demonstrated the temperature-dependent sheet resistance study for $V_2O_3$ thin films having a thickness of 7, 14, 28, and 56 nm. Electrical conductivity determined at room temperature increased from 400 S/cm (7 nm thin film) to 2122 S/cm (58 nm thin film), which is excellent from a *p*-type TCO perspective. They exhibited very low Hall mobility values, varied between 0.14-0.24 $cm^2$/V-s, but as-deposited thin films acquired great hole concentration of 1.77-5.28×$10^{22}$ $cm^{-3}$, which are highest among the previously reported *p*-type TCOs. On the contrary, as expected, optical transparency was decreased from 78% (7 nm thin film) to 40% (58 nm thin film) of $V_2O_3$ thin films, as depicted in figure 14. However, this better performance *p*-type TCO material can be used for a variety of optoelectronic applications.

Recently, Ainabayev et al. (2024) reported the successful synthesis of high-quality epitaxial *p*-type $V_2O_3$ thin films, demonstrating excellent electrical performance with measurable mobility and high carrier concentration [170]. The films displayed conductivity ranging from 115 to 1079 S/cm and optical transparency varying between 32 to 65% in the visible region. Mobility limitations in thinner films were attributed to their nanosized granular structure and the presence of two preferred growth orientations. Particularly, the 60 nm thick $V_2O_3$ film exhibited a highly competitive transparency-conductivity figure of merit (~5700 µS) compared to current standards in the field.

Apart from the above materials, there are various oxide materials explored as possible effective *p*-type TCOs using various deposition techniques and parameters. Their optoelectronic results are highlighted in table 7 [171–174]. Among these, $La_{1-x}Sr_xVO_3$ materials deposited using PLD technique demonstrated excellent optoelectronic properties and figure of merit (~6800 µS), which can be suitable candidate for *p*-type TCO applications.

### 1.5 Considerations for developing *p*-type TCOs:

In *n*-type TCOs, CBM mostly consists of metal s-orbitals, which are spatially distributed. Hence, high conduction can be achieved due to the low effective mass of electrons and high-density electron doping. That's why most *n*-type TCOs exhibit an enormous amount of electrical conductivity. But considering the *p*-type oxides, VBM is composed of oxygen 2p orbitals which are highly localized and lead to a higher effective mass of holes. The major concern in *p*-type TCOs is the strong localization of positive holes at valence band maximum (VBM) in oxide materials. This localization is due to the high electronegative nature of oxygen and its energy level lying far lower than the VBM of metallic atoms [175]. After reviewing the literature, we have defined some key points which need to be considered for developing a *p*-type TCOs:

**(i)** Choose a material with optical band gap energy higher than 3 eV to ensure high transparency in the visible region.

**(ii)** Choose a material whose host cation has equal or higher energy-filled orbital as compared to oxygen 2p orbital.

**(iii)** Select a suitable countercation with high energy outer-shell orbital, which can destroy the robust localization around oxygen anions (CMVB technique) and disperse the valence band and make holes more mobile. Otherwise, there is a probability that the generation of excess charge carriers can roam around oxygen atoms and cannot drift inside the crystal lattice, even under the action applied electric field.

**(iv)** Earlier, it was assumed that host material cation or counter-cation used for substitutional doping should have a closed-shell electronic configuration to avoid coloration. But, with time, it was realized that intrinsic electronic transition due to partially filled d-orbitals didn't affect the optical transparency as much. Therefore, the element with partially or half-filled electronic configuration, such as transition metal oxides, can also be used as host material or dopant. Chromium oxide ($Cr_2O_3$) material is a prime example of that.

**(v)** The coordination number in a crystal structure of material also plays a vital role in improving electrical properties. Materials having tetrahedral coordination is beneficial for hole conductivity, as it can disturb the localization around oxygen anions [42]. In this coordination, oxide anions exhibit $sp^3$ hybridization. Hence, all eight electrons of oxygen anions are coordinated with four sigma bonds with adjacent cations, decreasing the non-bonding nature of oxygen anions and diminishing the localization at the valence band maximum (VBM). For instance, $Cu_2O$ material exhibits excellent *p*-type conductivity as they also acquired tetrahedral coordination, and Cu outer shell orbital energy is more than the oxygen 2p level. But, the optical band gap of 2.1 eV limits its use for TCO applications [176,177].

### 1.6 Factors affecting the properties of TCO thin films:

Apart from selecting TCO material, the concentration of dopants/defects, deposition techniques, and its parameters also equally affect TCO's optical and electrical properties in thin-film form. These parameters include deposition temperature, time, base pressure, background gas (such as argon, oxygen, nitrogen, etc.) pressure, target material, and nature of the substrate have influence on the optoelectronic properties of the TCO thin films. Hence, after the material selection, all these parameters alongwith techniques also need to be optimized for achieving the effective performance of TCO.

### 1.6.1 Nature of substrate:

Types of substrates used for the deposition of TCO thin films can affect its optical and electrical properties. Different substrates appear to have great influence on the crystalline nature (whether single or polycrystalline) and sheet resistance of as-deposited thin films. The crystallinity of TCO thin film is a very crucial factor for determining its properties. As it can affect TCOs properties in many ways, such as grain boundaries between the crystallites, which can lead to scattering of charge carriers at the interfaces and decrease mobility and hence, conductivity. Grain boundaries are also linked with thin film's surface roughness, increasing the scattering (diffuse reflectance) at the surface, and degrading its optical transparency.

### 1.6.2 Concentration of dopant:

As mentioned earlier in the introduction part that intrinsic material is mostly insulating in nature. Hence, appropriate dopants must be used for enhancing the electrical properties without much compromising with optical properties. Because introduction of dopant may result in the generation of excess charge carriers, causing rise in optical absorption and decrease the transparency. Considering the electrical properties, an increase in dopant concentration improves thin films conductivity, but upto a certain limit. After that, various scattering mechanisms such as ionized impurity scattering, grain boundary scattering, and phonon scattering can negatively affect the electrical properties. Besides, the amount of dopant above the solubility limit results in the formation of secondary phase, affecting both optical and electrical properties. Hence, the optimized amount of dopant into host material must be identified.

### 1.6.3 Defects contribution:

Materials (whether in bulk or thin-film form) demonstrate *n*-type or *p*-type conductivity without any doping. The sources of this electrical conductivity are their native defects present within the materials. The existence of defects and their types can be determined using density functional theory (DFT) calculations. DFT calculations identify the formation energy of various defects, whether in the form of equilibrium concentrations, the stability of different oxidation states or their intrinsic electronic transitions [178]. Hence, it is necessary to identify the contribution of defects to electrical conductivity of materials, so that their deposition parameters can be optimized during thin film growth.

**1.6.4 Deposition techniques:**

There are varieties of thin film deposition techniques that can be used to grow TCO thin films. It all depends upon the purpose for which thin films are being prepared. Chemical methods involve the deposition of volatile precursor material on the substrate, whereas physical methods comprise the ejection or evaporation of solid from the target material. But chemical methods offer several advantages due to their simplicity, cost, homogeneity, and no need for a vacuum system. However, physical methods result in more smooth and epitaxial crystalline thin films, beneficial for TCO's optoelectronic performance in thin-film form.

**1.6.5 Deposition temperature and time:**

At what temperature, deposition of thin films being processed is also important and must be optimized. Because it decides the crystallinity of the as-deposited thin films and can also affect the electrical properties. Apart from that, deposition temperature can also influence the formation of phase. As there are various reports in the literature where the variation of deposition temperature leads to secondary or unwanted phase [154]. Similarly, deposition time also has a respective influence on the properties of TCO thin films.

**1.6.6 Background gas pressure:**

The optoelectronic properties of TCO can be tailored by introducing background gas in the deposition chamber. The background gases such as argon, nitrogen, or oxygen, and their partial pressure affects *n*-type TCO and *p*-type TCO properties differently. Like in *p*-type TCO, usually, oxygen gas is used to recover the oxygen vacancies, which is beneficial for optical properties. Oxygen at high partial pressure may have the probability of entering the interstitial position, resulting in the generation of extra charge carriers and improved electrical properties. Figure 15 illustrates the structural differences between a stoichiometric $ABO_2$ crystal and a non-stoichiometric $ABO_2$ crystal with "excess" oxygen atoms. These structural differences can significantly affect the physical and chemical properties of the material. The excess oxygen atoms can introduce defects and alter the electronic structure, leading to changes in the material's behavior and potential applications [179]. In various reports, post-annealing of thin films was performed at high oxygen partial pressure for several hours and resulted in a remarkable improvement in hole conductivity. Hence, the use of background gas with optimized partial pressure is necessary to enhance TCO thin films performance.

**1.7 Summary and Remarks:**

**I.** We reviewed the basic materials physics of both conventional and newly discovered *p*-type TCO materials in this study. There have been reports on the utilization of non-stoichiometry and doping for *p*-type TCO thin films using various techniques. Newer and better *p*-TCO thin films can be produced with a deeper comprehension of the defect chemistry and the function of dopants in raising the hole conductivity of these materials. Increasing delocalization of the valence-band edge from oxide ions to metallic cations was suggested to be achieved by adding covalency to the typically ionic bonds between cations and oxygen ions. The importance of crystal structure in determining the production of *p*-type TCOs was emphasized.

**II.** Despite significant progress, the conductivity of *p*-TCO thin films is still very less as compared to their *n*-TCO counterparts, presenting a critical challenge for the advancement of *p*-TCO technology. Various techniques such as CMVB and co-doping etc. have been proposed to develop new *p*-TCO thin films with better electrical and optical properties. Many deposition techniques have been adopted with widely varying growth parameters. For *p*-TCO technology, the biggest obstacle to producing high-performance active devices appropriate for **"Invisible Electronics"** is to increase the conductivity of *p*-TCO thin films without compromising their visible transmittance.

**III.** However, encouraging developments in material synthesis methods provides some optimism, as they may simplify fabrication procedures and accelerate the creation of transparent diodes. The development of high mobility tunnel field effect transistors (TFETs) and the reappearance of *p*-type ZnO thin films are significant directions for *p*-TCO technology, with the potential to significantly transform the transparent electronics market.

**IV.** Lastly, it is worth discussing that the objective of the applications for which *p*-type TCOs are intended should be closely tied to the selection criteria. In optoelectronic devices, low hole concentration and high mobility are desired, such as transparent TFT and *p*–*n* diodes. A high hole conductivity and transparency for UV to IR regions become more crucial for usage as the hole transport layer in solar cells. The materials do not necessarily need to be highly mobile and transparent to visible light to meet these criteria. Transition metal oxides doped with holes appear to satisfy this condition since these kinds of materials can be modified to obtain a high concentration of hole carriers at the d orbitals.

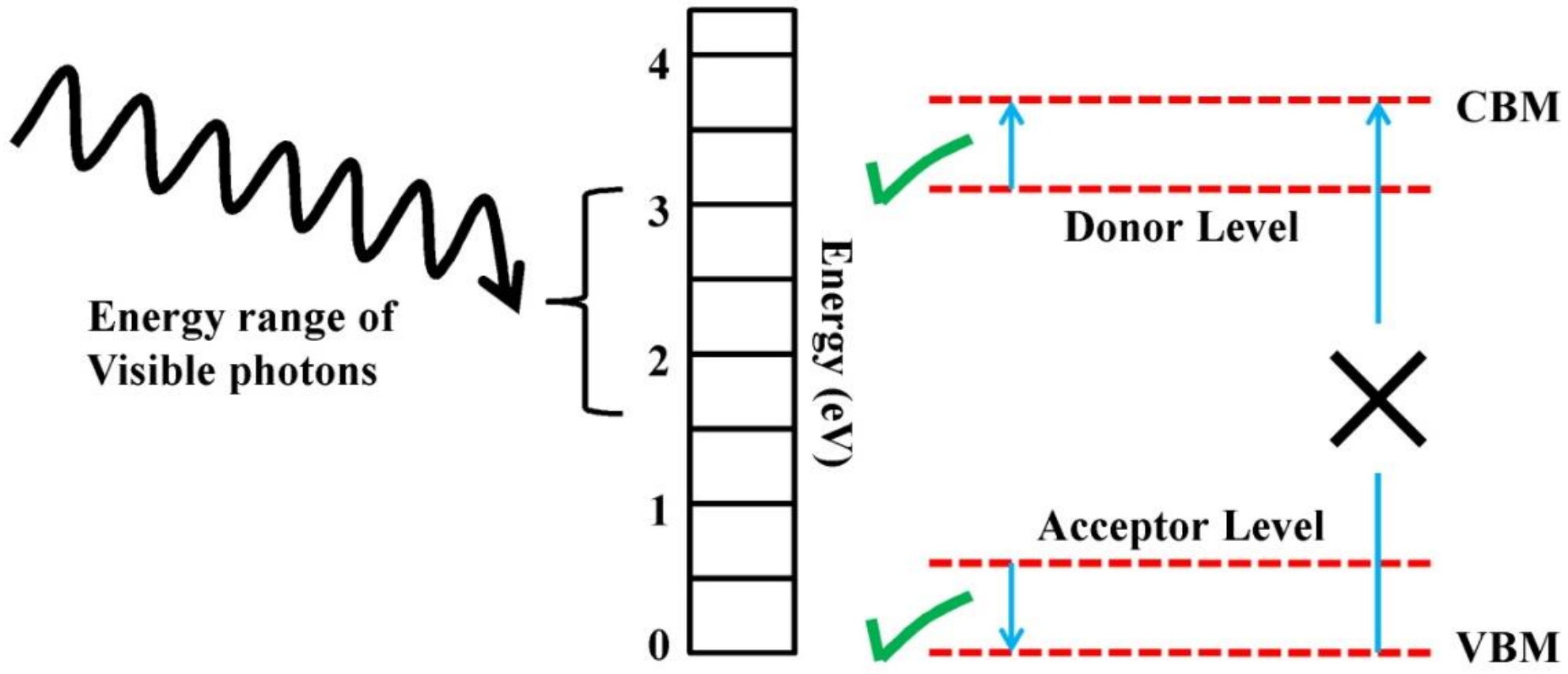

**Figure 1.** Band gap engineering of TCO material. Visible photons don't have enough energy to excite electrons directly from the valence band to conduction band. Still, they have sufficient energy to excite electrons from donor level to conduction band (in *n*-type TCO) or holes from the acceptor level to valence band (in *p*-type TCO). The symbol (✓) and (×) represents the allowed and forbidden transition.

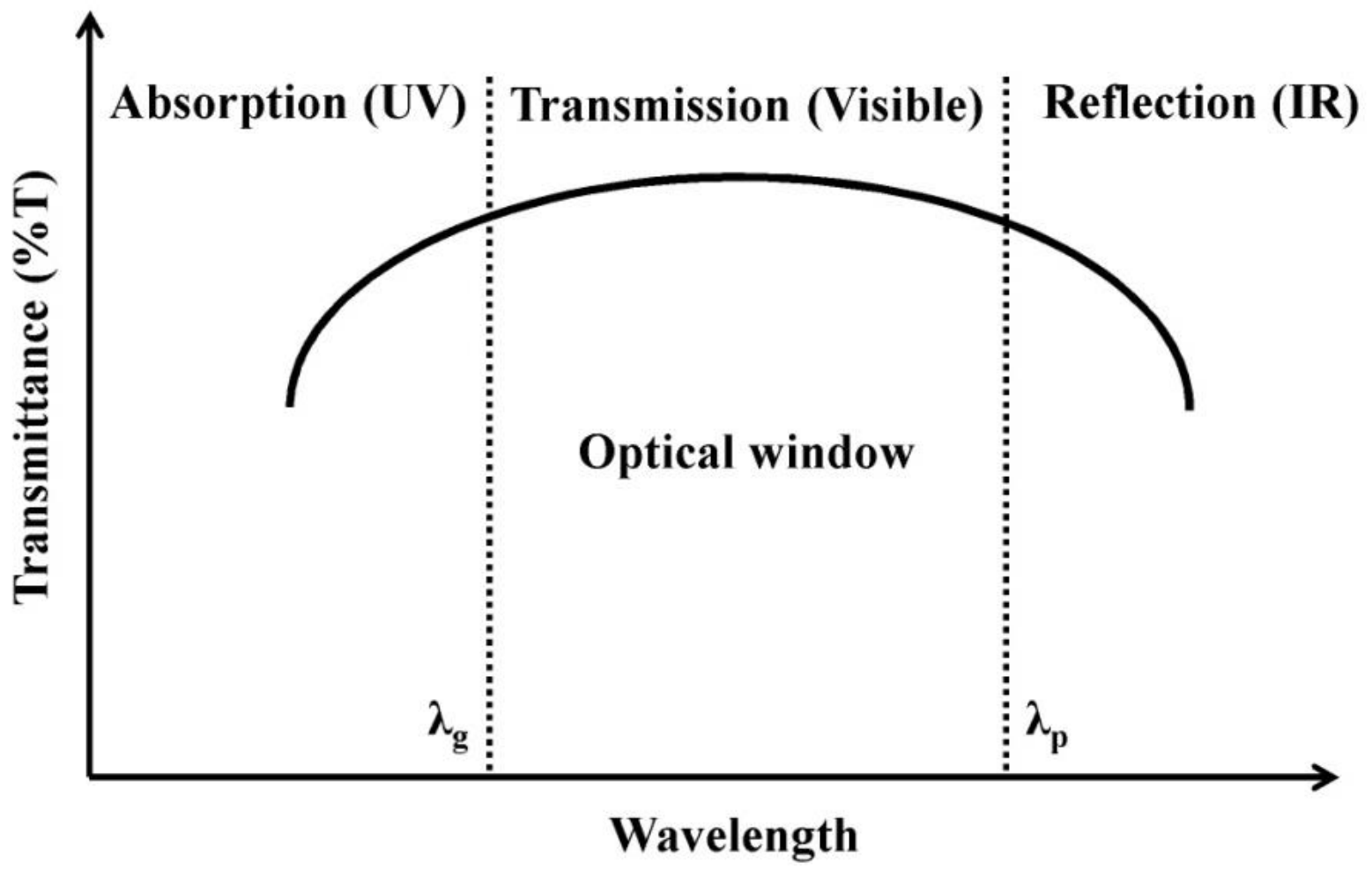

**Figure 2.** Transmittance spectrum of TCO material.

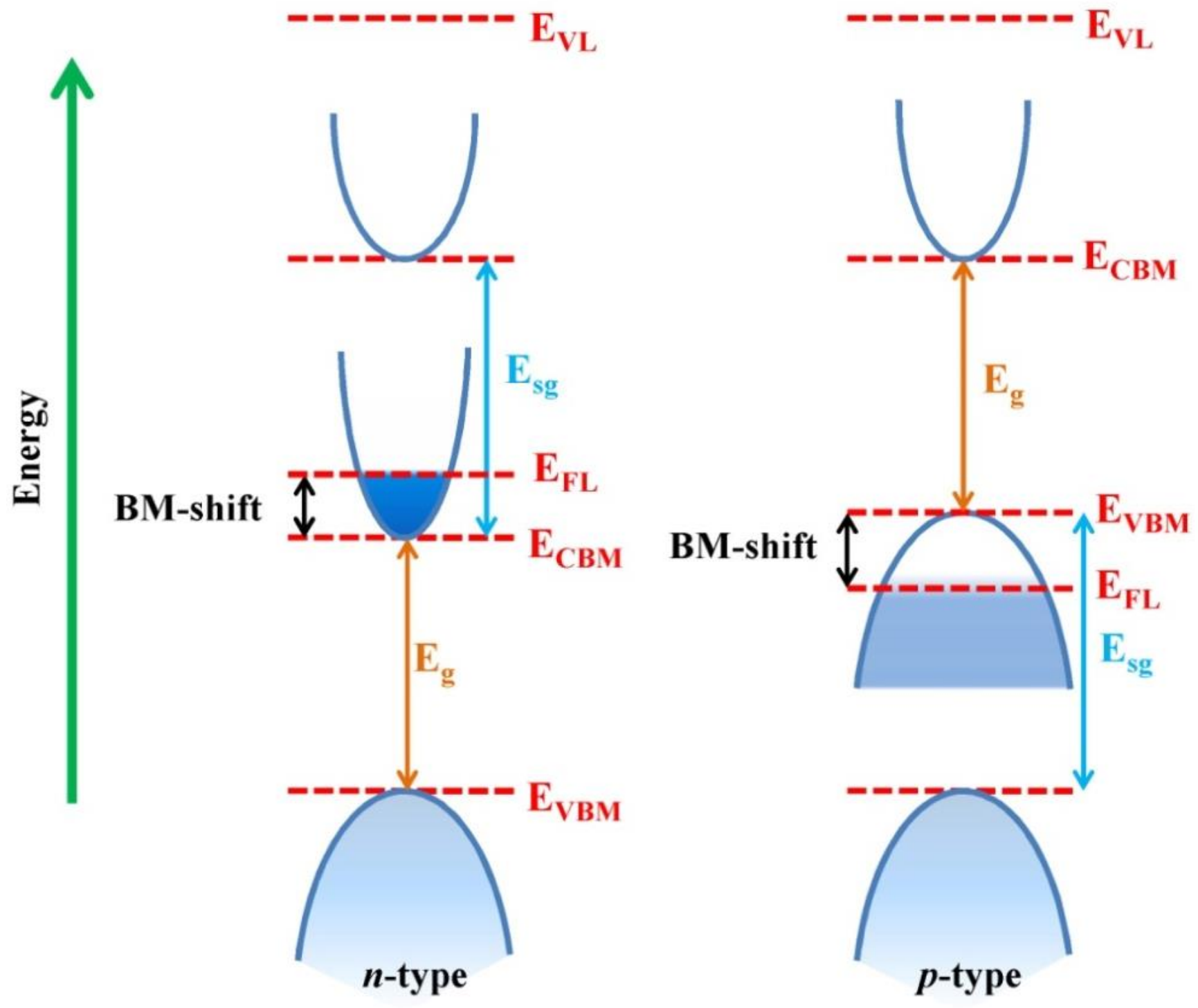

**Figure 3.** Band structure overview of *n*-type and *p*-type material with degenerate doping, illustrating the BM-shift, Fermi-level ($E_{FL}$), vacuum level ($E_{VL}$), fundamental band gap ($E_g$), and secondary band gap ($E_{sg}$). The dark blue shading portion represents the occupied energy states, while empty one is unoccupied energy levels.

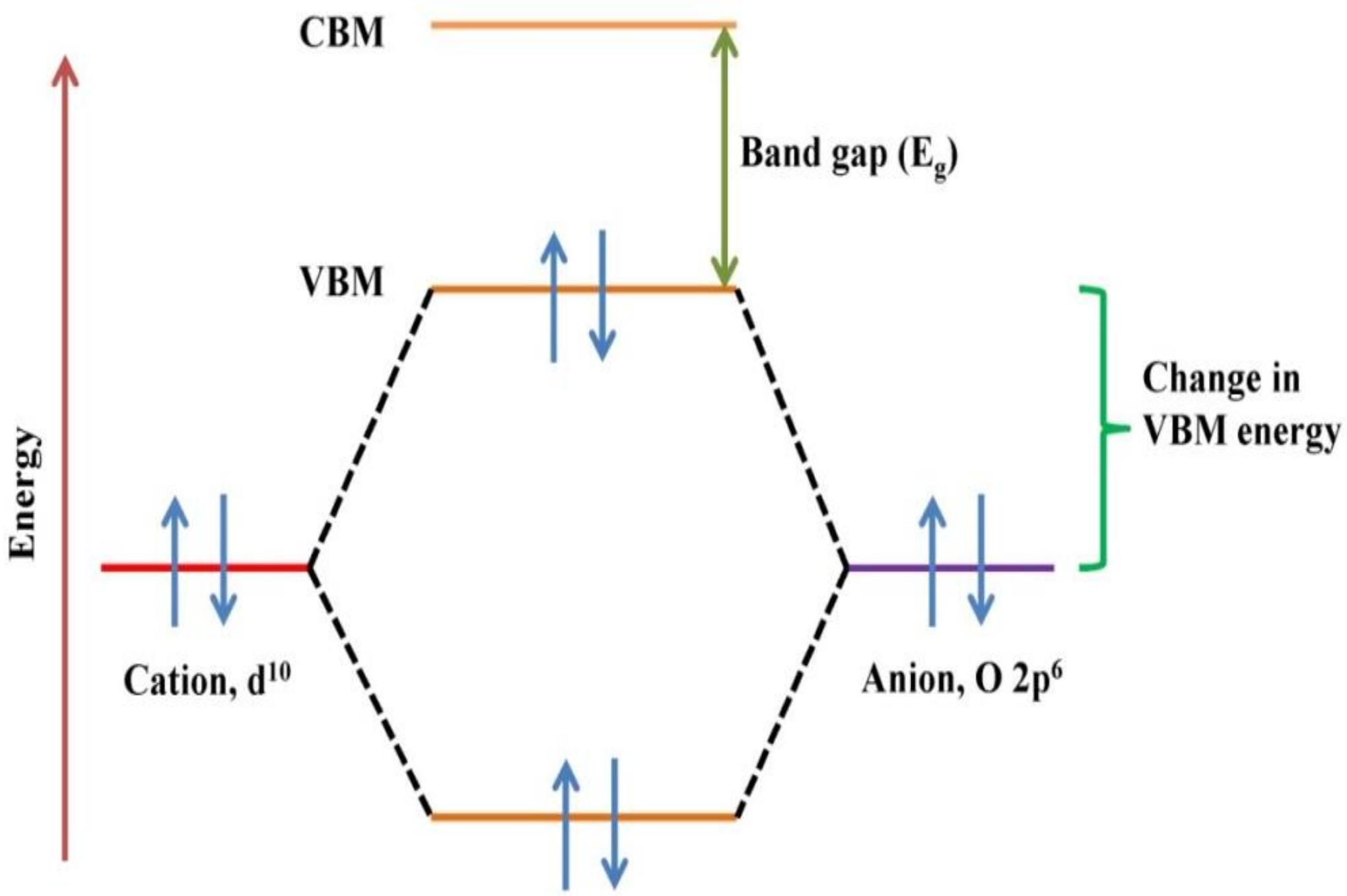

**Figure 4.** Schematic overview of CMVB technique [42]. Energy levels are not to scale.

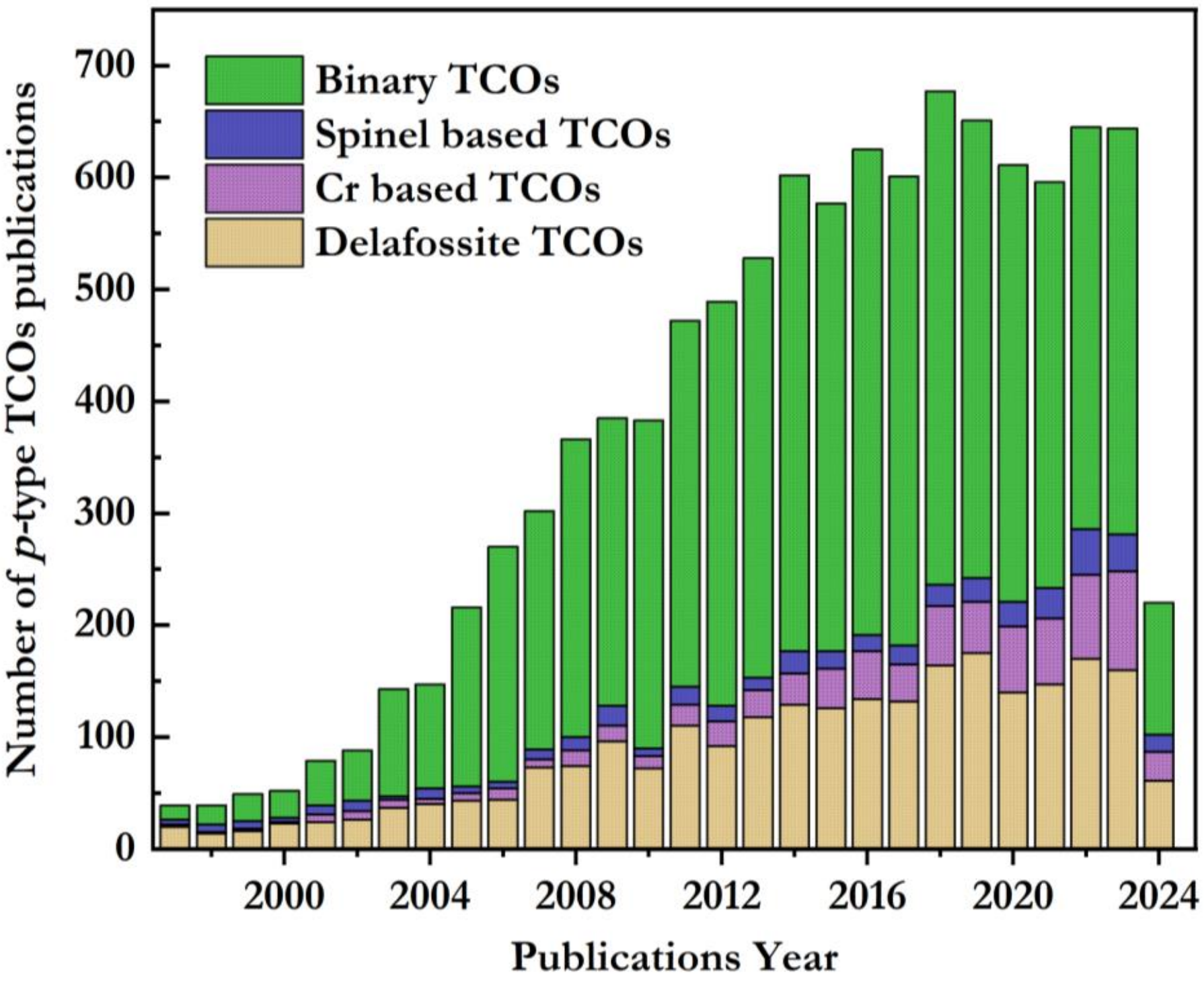

**Figure 5.** Evolution of publications on *p*-type TCO materials.

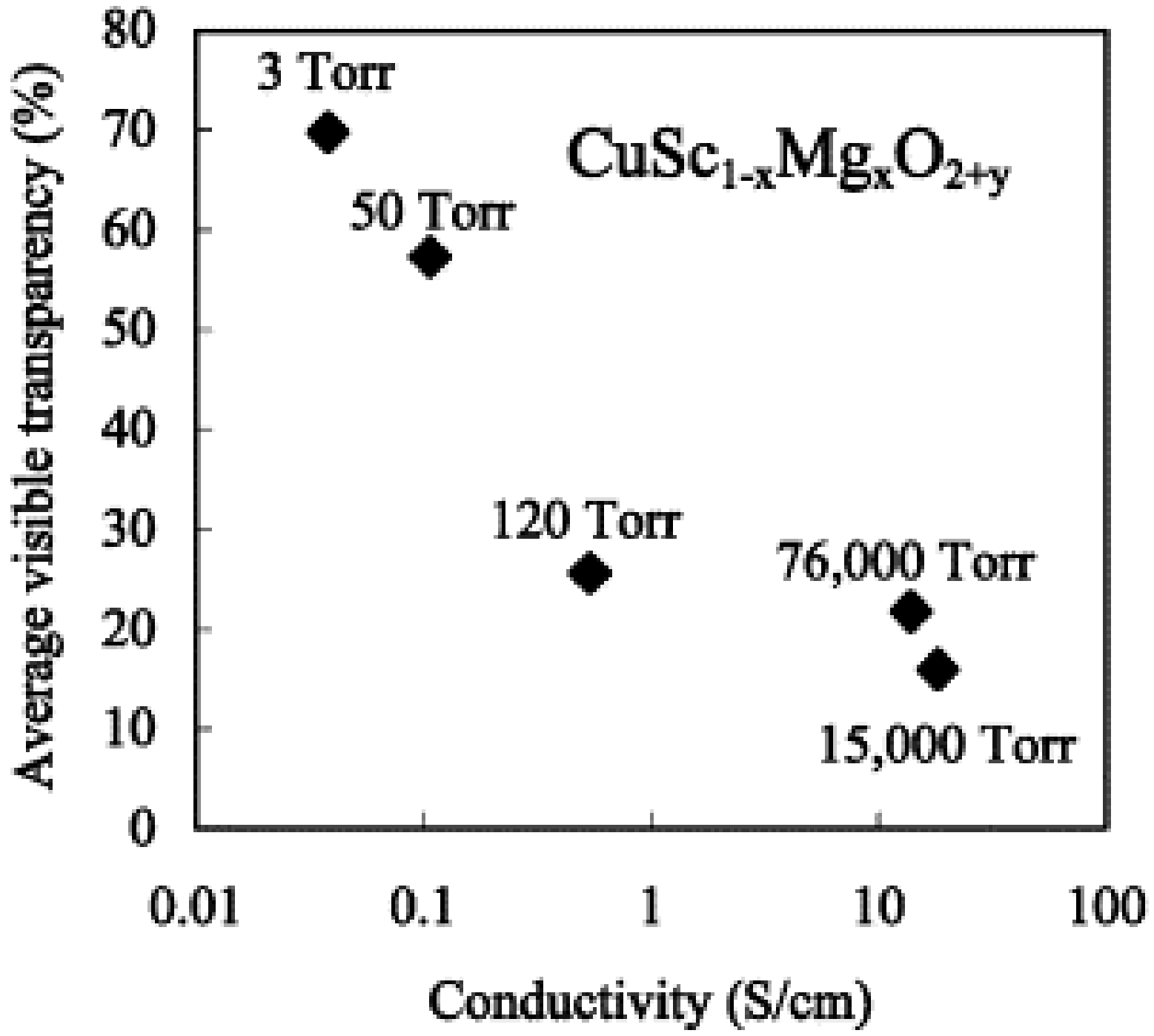

**Figure 6.** Variation in electrical conductivity of Mg-doped $CuScO_{2+y}$ thin film as a function of oxygen intercalation pressures [67].

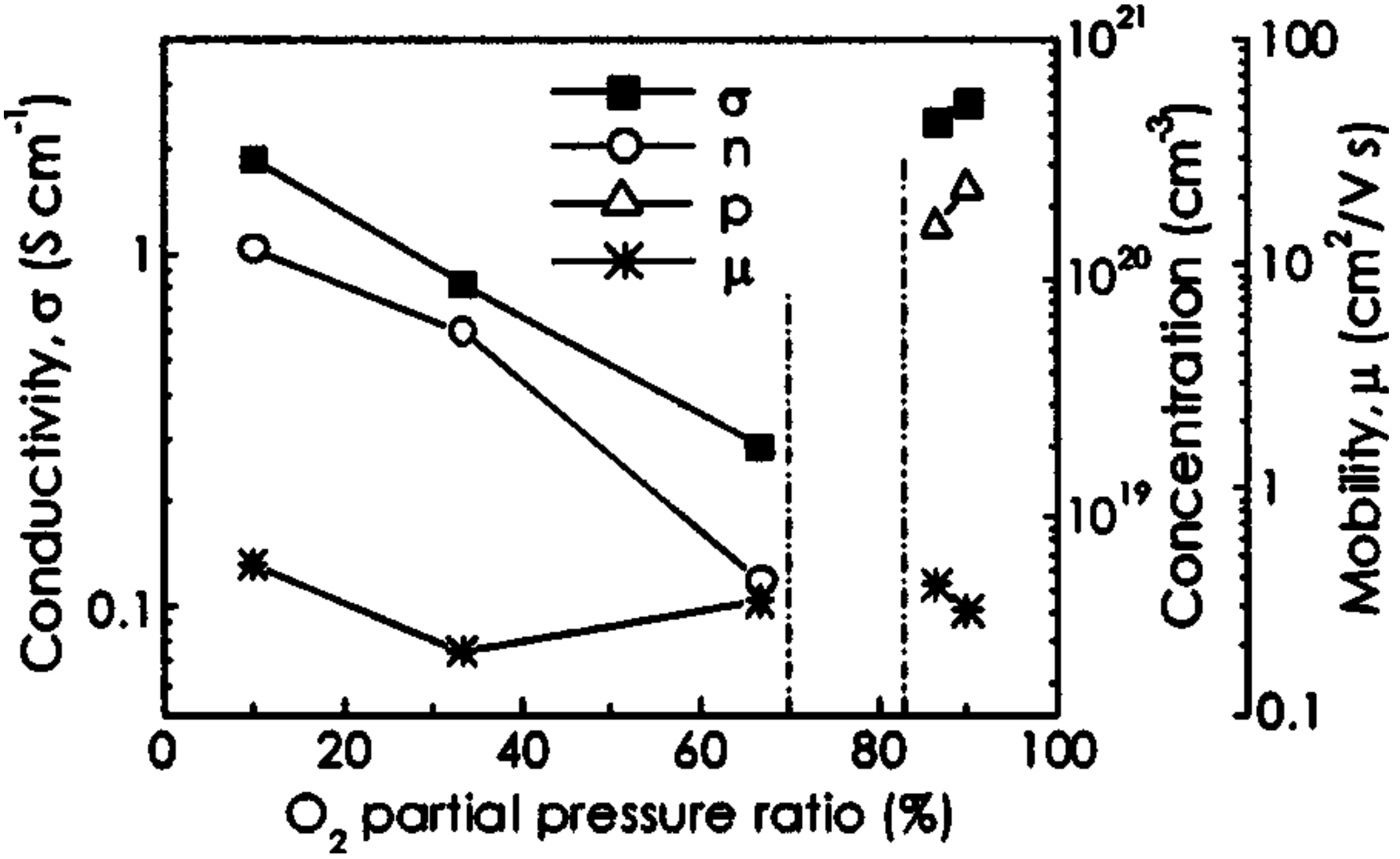

**Figure 7.** Variation in electrical conductivity (σ), carrier concentration (n) and mobility (μ) of $ZnCo_2O_4$ thin films as a function of oxygen partial pressure ratio [86].

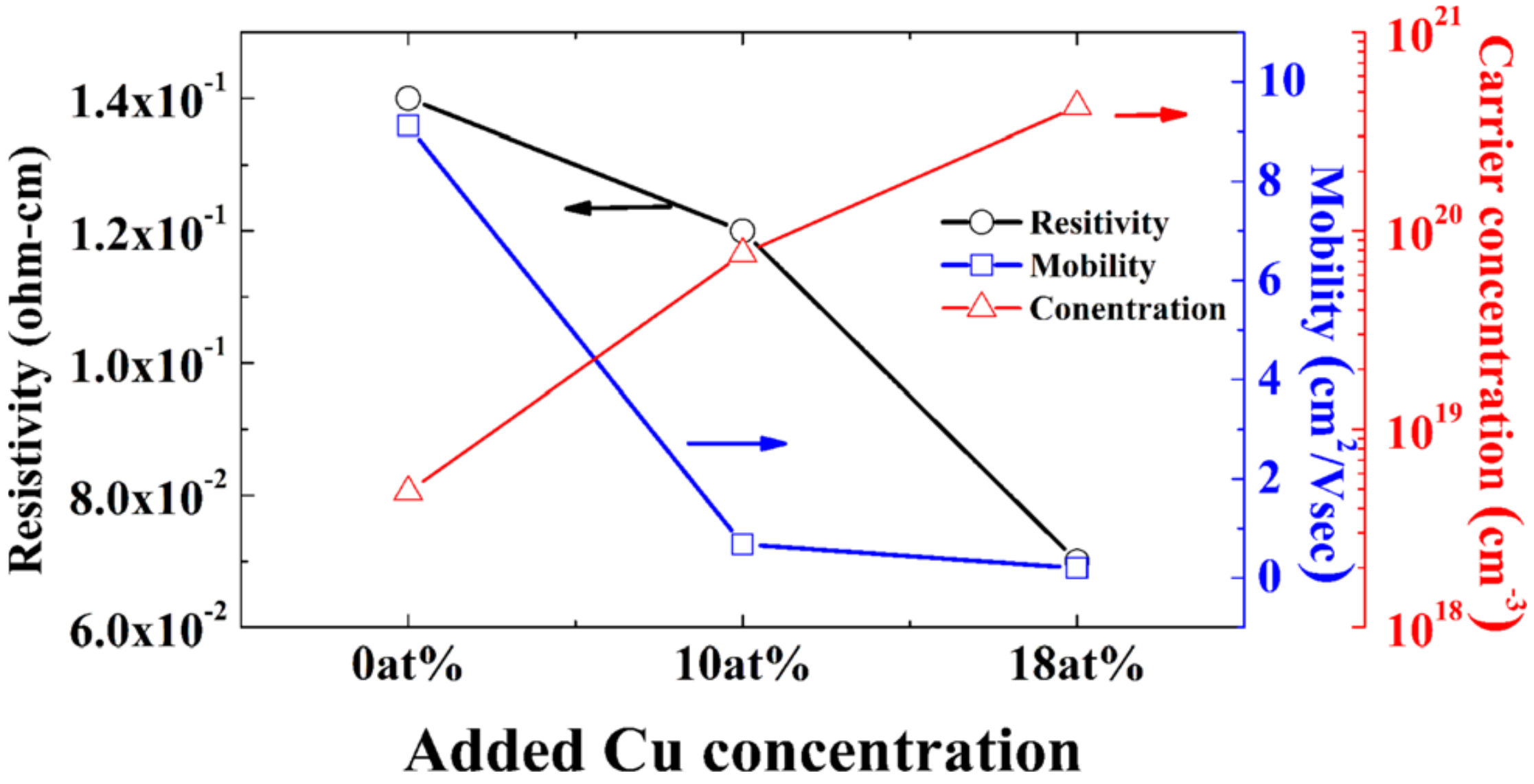

**Figure 8.** Resistivity, mobility, and carrier concentration of NiO thin films with varying Cu-content [98].

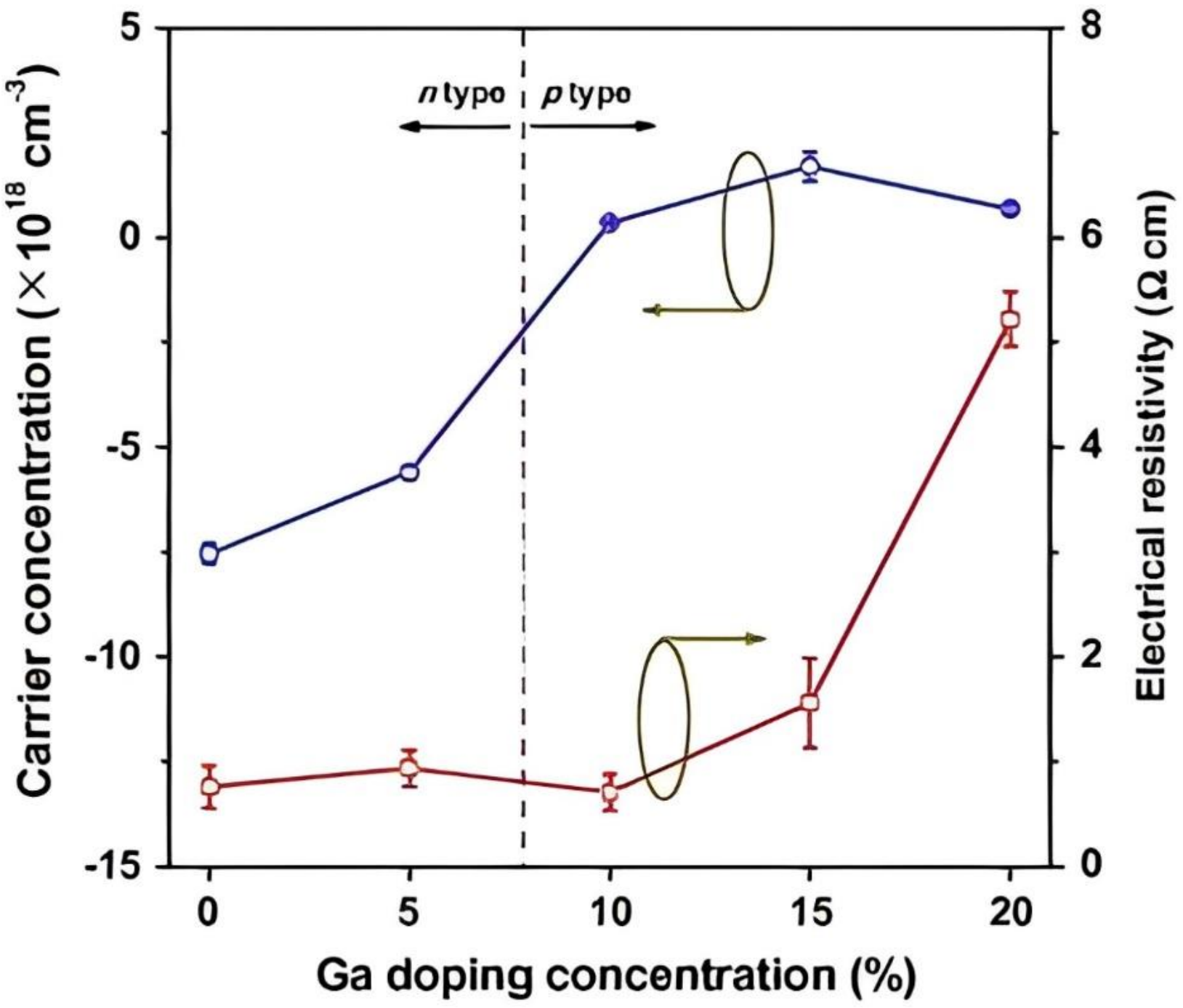

**Figure 9.** Changes in electrical resistivity, carrier type and carrier concentration of Ga-doped $SnO_2$ thin films [62].

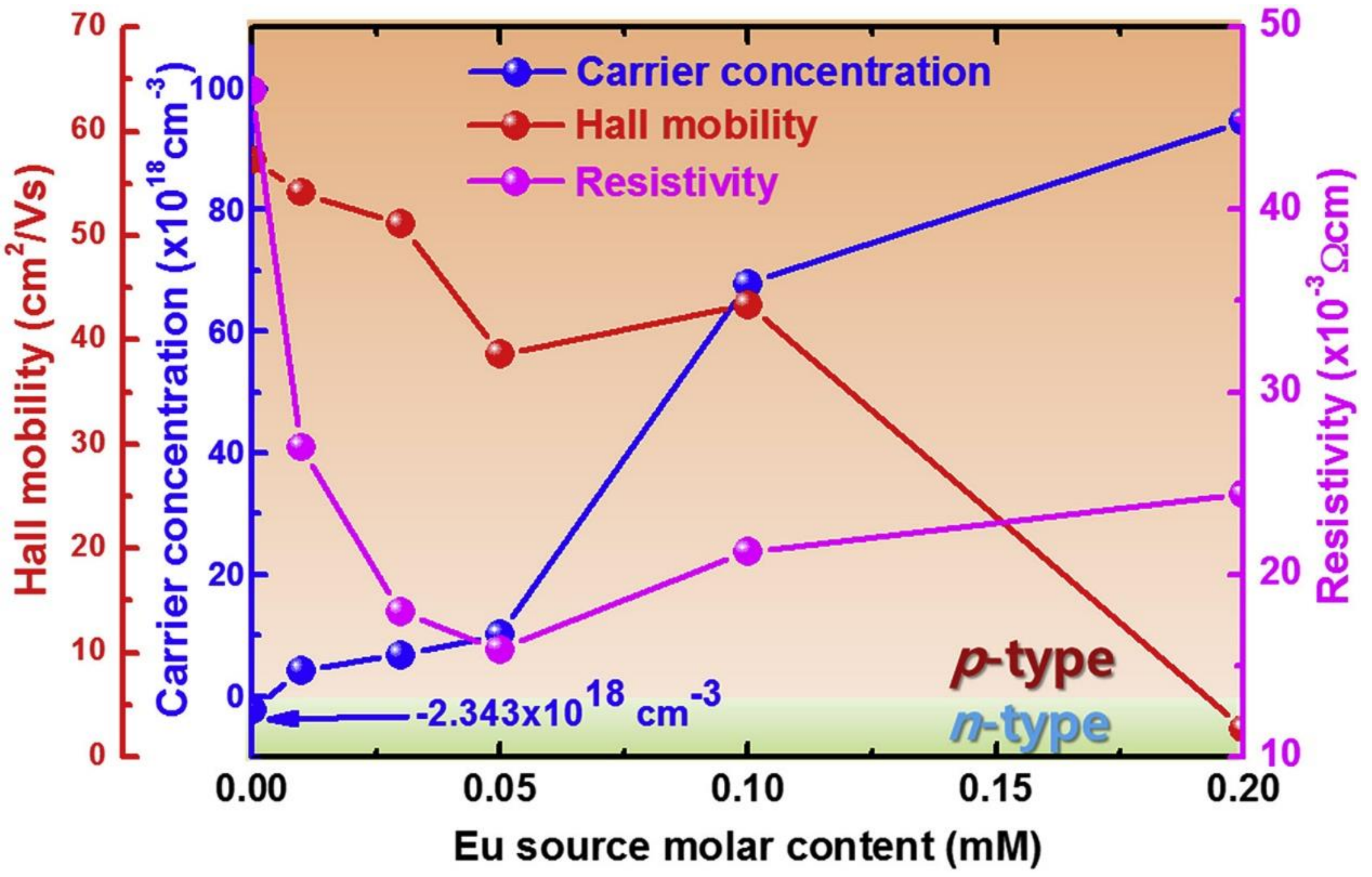

**Figure 10.** Changes in resistivity (ρ), Hall mobility (μ), and carrier concentrations of Eu-doped $SnO_2$ thin films with varying Eu molar concentration [130].

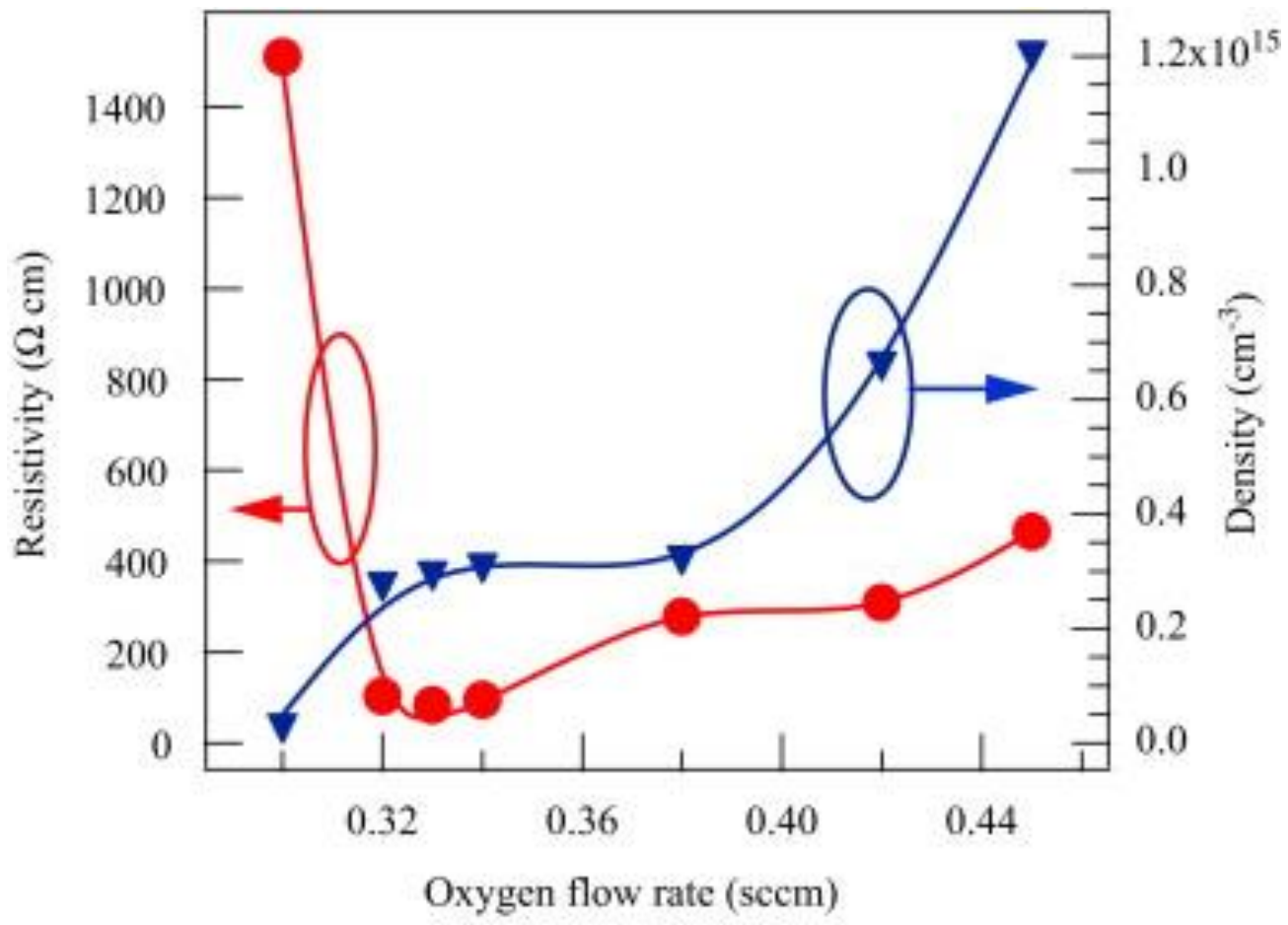

(a)

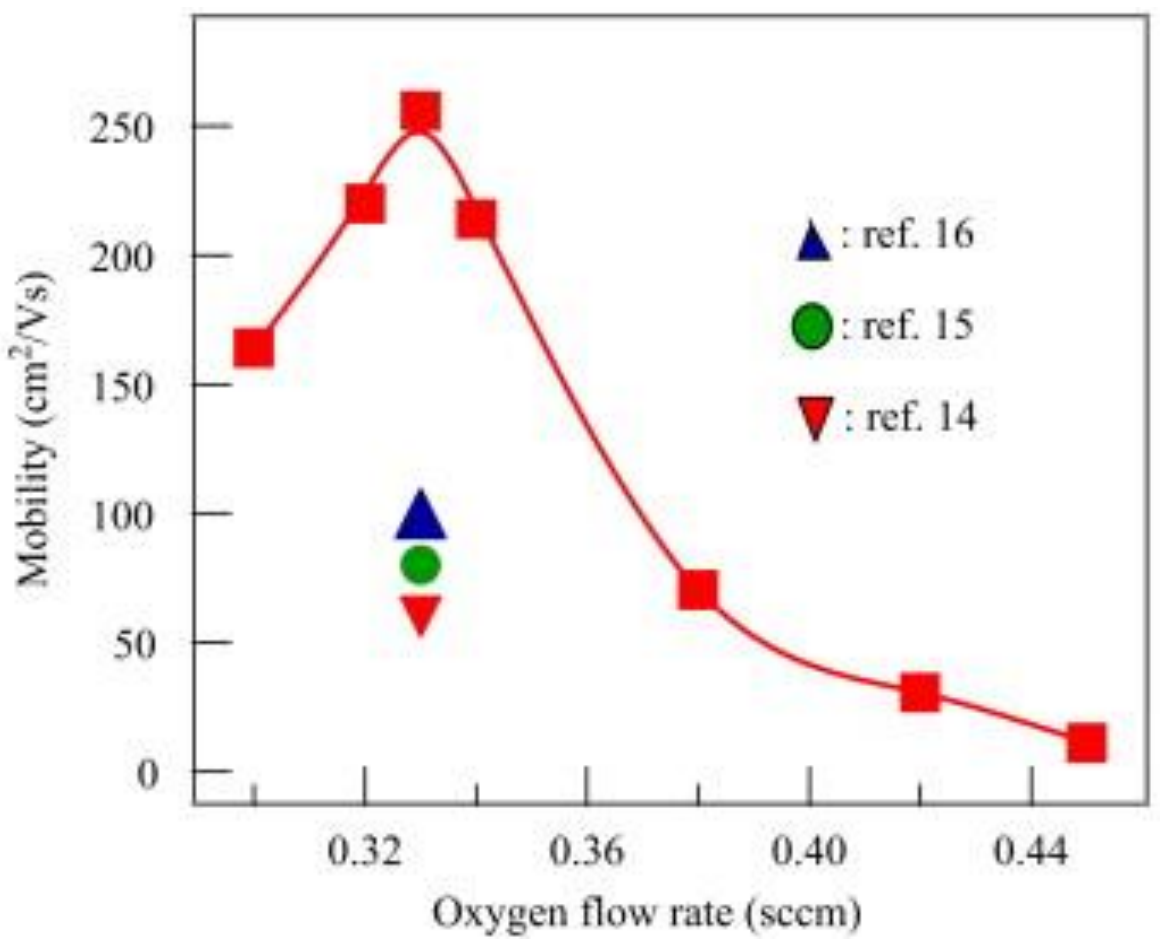

(b)

**Figure 11.** (a) Resistivity and hole density and (b) mobility of $Cu_2O$ thin films with LTB-$Cu_2O$ as a function of oxygen flow rates.

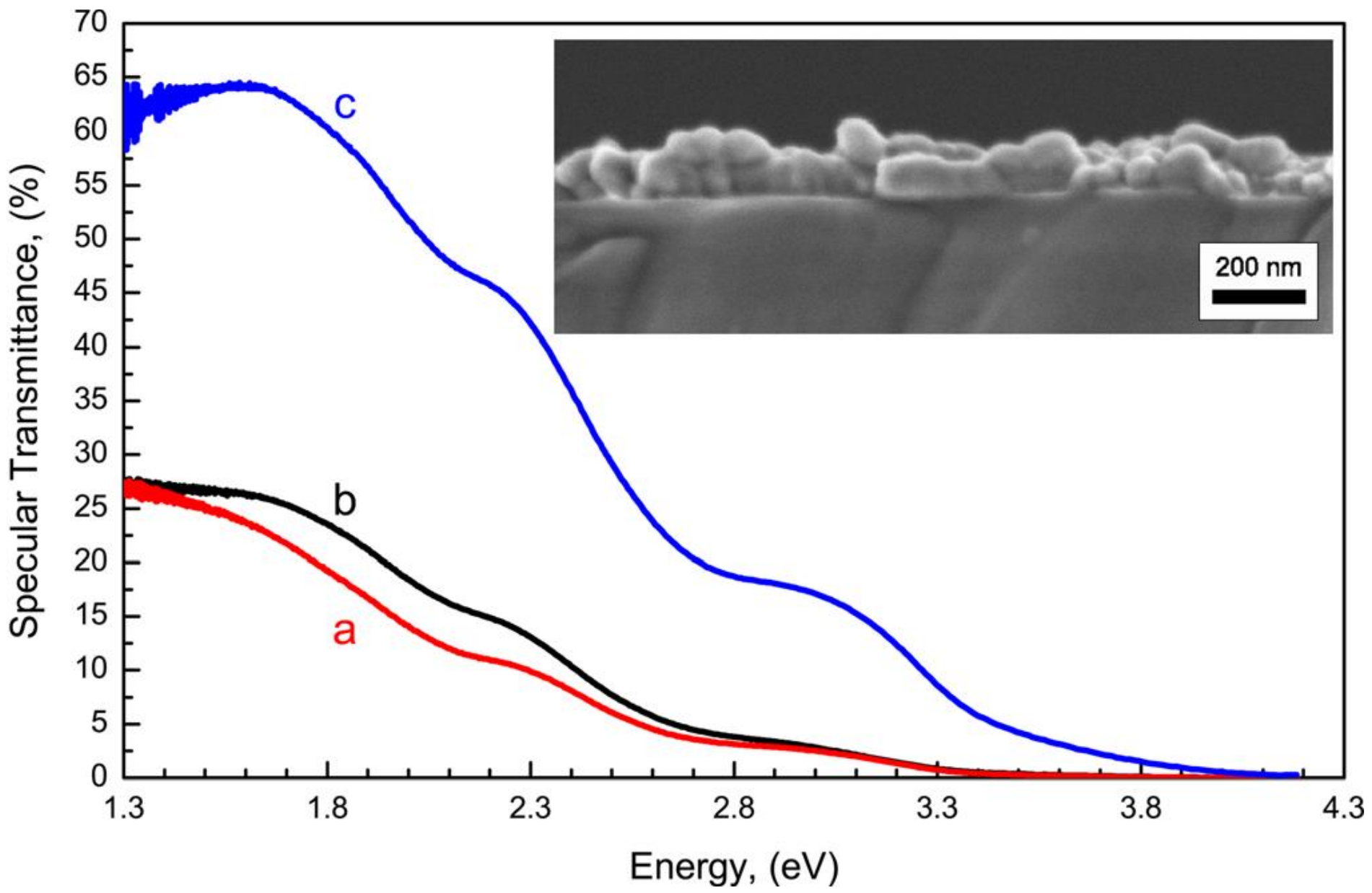

**Figure 12.** Optical transmission spectra of (a) $Cr_2O_3$, (b) Mg:$Cr_2O_3$, and (c) Mg,N:$Cr_2O_3$ thin films deposited with precursor ratios of Cr:Mg 9:1 and Cr:N 1:4. The inset shows a cross-section SEM image to measure film thickness [152].

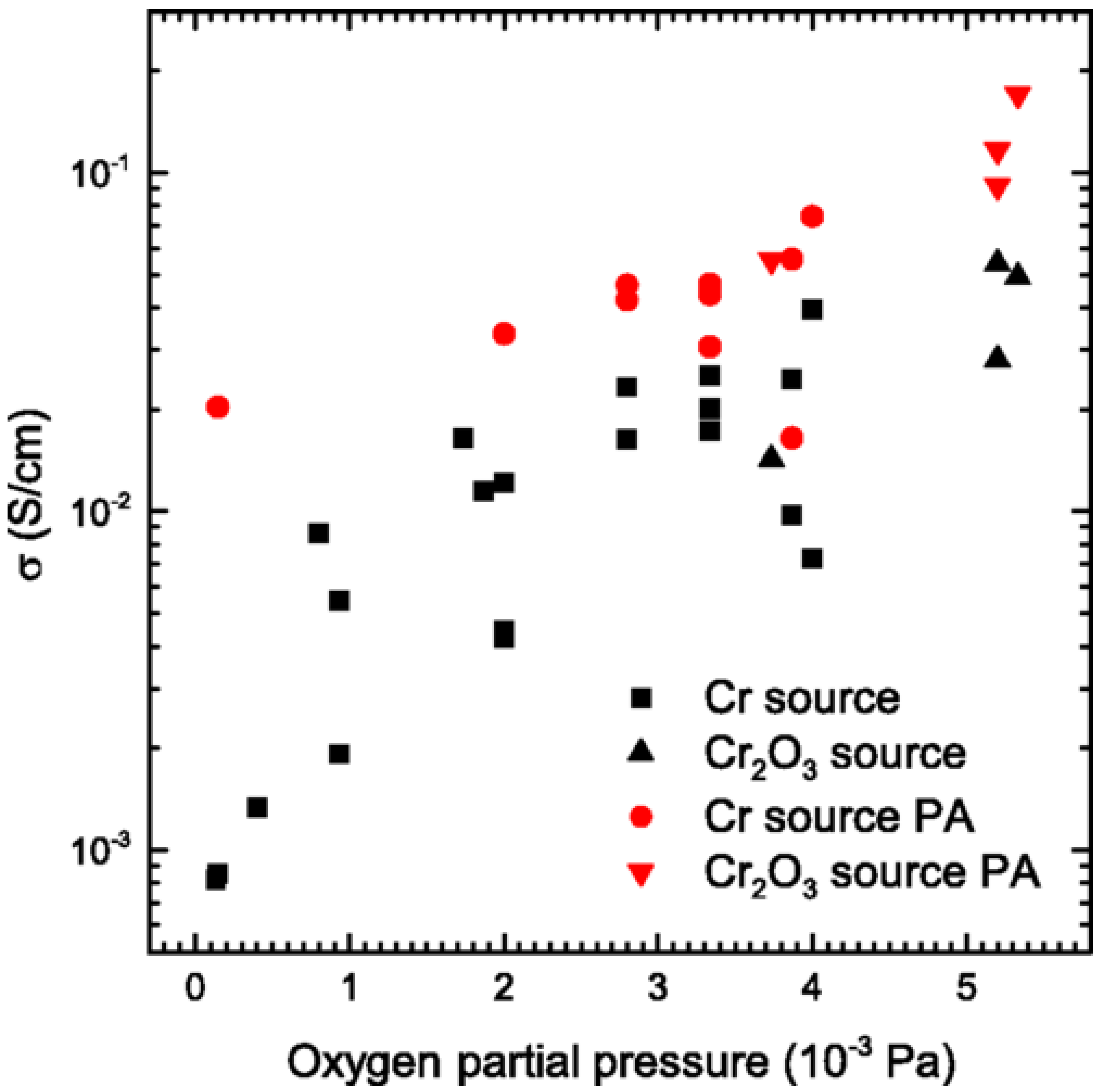

**Figure 13.** The room temperature electrical conductivity of $Cr_2O_3$ thin films (deposited from various sources) measured at different post-annealing oxygen pressures [153].

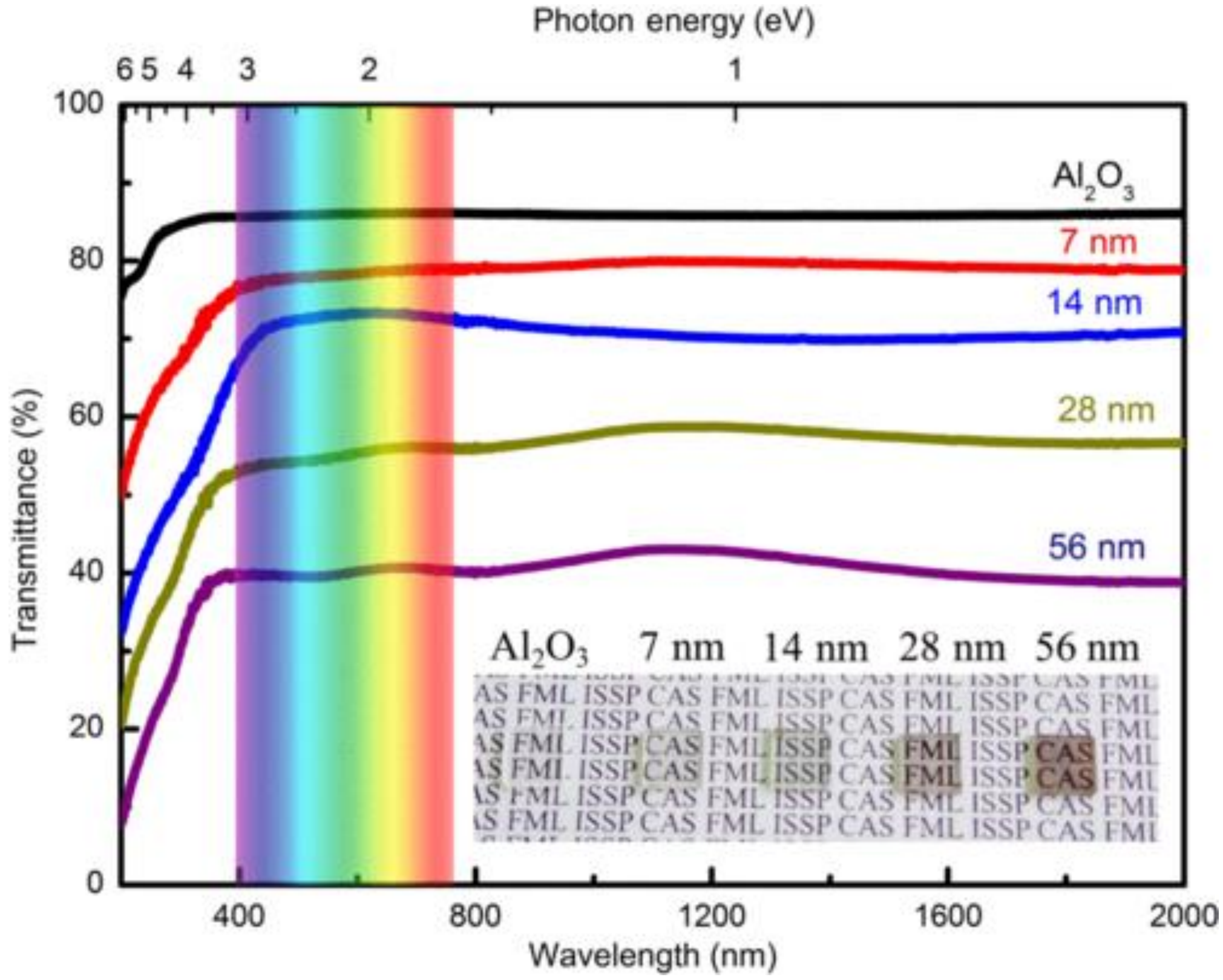

**Figure 14.** Optical transmittance spectra of $Al_2O_3$ substrate and $V_2O_3$ thin films of varying thicknesses. The inset illustrates the images of the $Al_2O_3$ substrate and $V_2O_3$ thin films with different thicknesses [169].

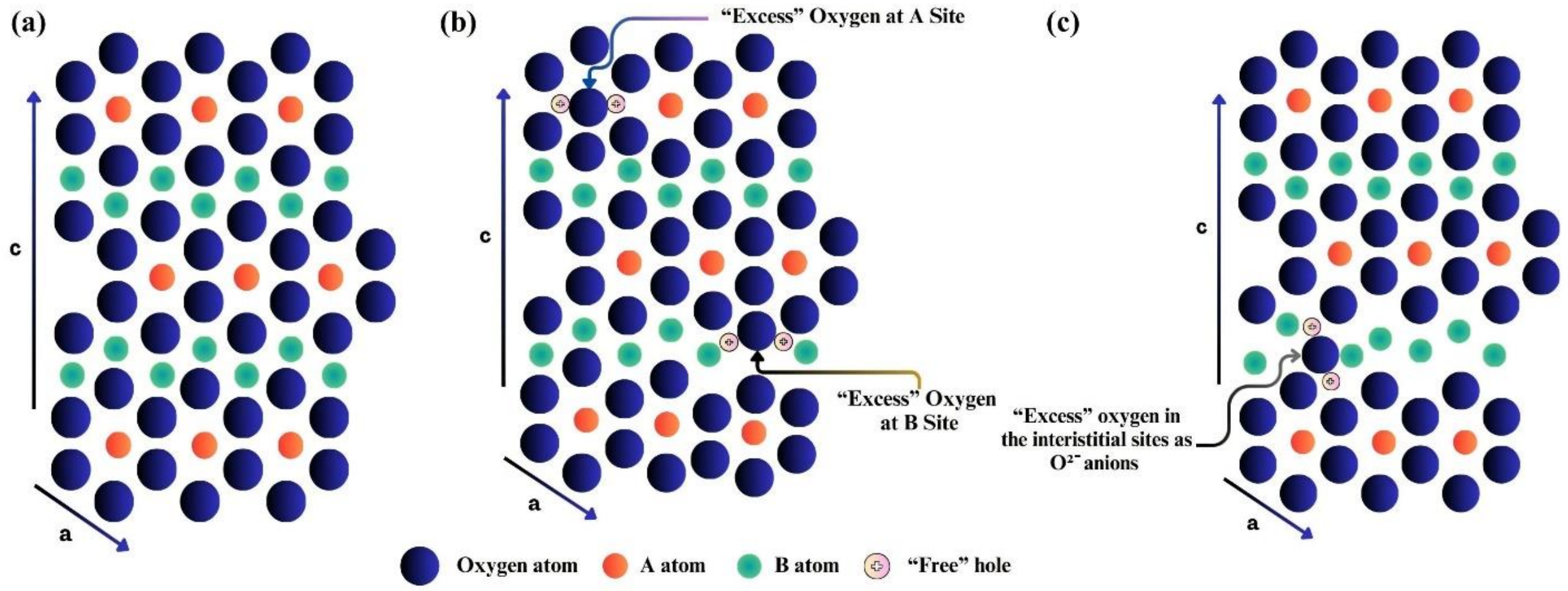

**Figure 15.** $ABO_2$ crystal lattice: (a) stoichiometric structure. (b & c) non-stoichiometric with excess oxygen: (b) occupying A and B cation sites (creates four holes, *p*-type), (c) occupying interstitial sites (forms $O^{2-}$, creates two holes, *p*-type).

**Table 1:** Cu-based materials as *p*-type TCO and their respective optoelectronic parameters.

| Material | Deposition technique used | Thickness (nm) | Substrate used | Average optical Transmittance (%) | Optical band gap (eV) | Electrical conductivity (S/cm) | Carrier conc (/cm$^3$) | Hall Mobility (cm$^2$/V-s) | Reference |
|---|---|---|---|---|---|---|---|---|---|
| **$CuAlO_2$** | PLD | 500 | Sapphire | 70 | 3.5 | 1 | $1.3 \times 10^{17}$ | 10.4 | [42] |
| **$CuGaO_2$** | PLD | 500 | Sapphire | 80 | 3.6 | 0.063 | $1.7 \times 10^{18}$ | 0.23 | [49] |
| **Sn: $CuInO_2$** | PLD | 280 | Sapphire | < 50 | 3.9 | $3.8\times10^{-3}$ | - | - | [46] |
| **Ca: $CuInO_2$** | PLD | 170 | Sapphire | < 50 | ~3.9 | $2.83\times10^{-3}$ | - | - | [46] |
| **$CuScO_{2+x}$** | Sputtering | 110 | Silica | 40 | 3.3 | 15-30 | - | - | [47] |
| **$CuY_{1-x}Ca_xO_2$** | Thermal co-evaporation | 240 | MgO | 40-50 | 3.5 | 1 | - | - | [48] |
| **$CuCr_{1-x}Mg_xO_2$** | RF-sputtering | 250 | Quartz | 30 | 3.1 | 220 | - | - | [45] |
| **Cu-deficient $CuCrO_2$** | spray pyrolysis | 80 | Glass | 55 | 3 | 1-12 | - | 0.006 | [70] |
| **Mg, N: $CuCrO_2$** | RF-sputtering | 150 | Quartz | 69.1 | 3.52 | 277.7 | $1.18\times10^{21}$ | 0.006 | [71] |
| **Sn:$CuNi_{1-x}Sb_xO_2$** | RF-sputtering | 150-200 | Fused silica | 60 | 3.4 | 0.05 | - | - | [72] |
| **K-doped $SrCu_2O_2$** | PLD | 200 | $SiO_2$ glass | ~ 70 | 3.3 | $4.8\times10^{-2}$ | $6.13\times10^{17}$ | 0.46 | [50] |
| **$CuCrO_{2+x}$** | Combustion | 75 | ITO | 80-90 | 3.0 | 0.14 | $3.8 \times 10^{18}$ | 0.23 | [73] |
| **$CuCrO_2$** | CVD | 75-85 | Boro aluminosilicate glass | 60-65 | 3.2 | 50 | $9.3 \times 10^{18}$ -$1.5 \times 10^{16}$ | 0.65-5 | [74] |
| **$CuGaO_2$** | RF-sputtering | 200 | Fused quartz | 60-70 | 3.47 | 0.0167 | - | - | [75] |
| **$CuAlO_2$** | Magnetron Sputtering | 400 | Sapphire | 75 | 3.85 | 1.05 | $2.5 \times 10^{19}$ | 0.254 | [76] |

**Table 2:** Ag-based materials as *p*-type TCO and their respective optoelectronic parameters.

| Material | Deposition technique used | Thickness (nm) | Substrate used | Average optical Transmittance (%) | Optical band gap (eV) | Electrical conductivity (S/cm) | Carrier conc (/cm$^3$) | Hall Mobility | Reference |
|---|---|---|---|---|---|---|---|---|---|
| **$AgGaO_2$** | PLD | 200 | Sapphire | 50 | 4.12 | $3.2\times10^{-4}$ | - | - | [77] |
| **$AgCr_{1-x}Mg_xO_2$** | Spin Coating | 120 | Sapphire | 60 | 3.66 | $67.7\times10^{-3}$ | $1.25\times10^{17}$ | 50 | [79] |
| **Ag-deficient $AgInGaO_2$** | Reactive Evaporation | - | Glass substrate | 25 | 3.77 | 61 | $2.2\times10^{18}$ | 0.017 | [80] |

**Table 3:** Spinel materials as *p*-type TCO and their respective optoelectronic parameters.

| Material | Deposition technique used | Thickness (nm) | Substrate used | Average optical Transmittance (%) | Optical band gap (eV) | Electrical conductivity (S/cm) | Carrier conc ($/cm^3$) | Hall Mobility ($cm^2/V$-s) | Reference |
|---|---|---|---|---|---|---|---|---|---|
| **$NiCo_2O_4$** | Spin Casting | 100 | Sapphire | 90 | - | 16.7 | - | - | [81] |
| **$ZnRh_2O_4$** | RFMS | - | Silica glass | - | 2.1 | 0.7 | - | - | [82] |
| **$ZnCo_2O_4$** | RFMS | 400 | Silica glass | 55 | 2.63 | 1.2 | $2.81\times10^{20}$ | 0.2 | [86] |
| **$ZnIr_2O_4$** | PLD | 100-300 | Quartz and sapphire | 60 | 2.97 | 2.09-3.39 | - | - | [82] |
| **$ZnCo_2O_4$** | PLD | 100-200 | Sapphire | - | 2.3 | 21.8 | $2.6\times10^{16}$-$1.49\times10^{20}$ | 0.12-0.6 | [91] |
| **$ZnCo_2O_4$** | Spin coating | 120 | Glass | 55 | 3.95-3.99 | $0.4\times10^{-3}$-$0.6\times10^{-2}$ | $5.57\times10^{12}$-$4.29\times10^{14}$ | - | [89] |
| **$ZnGa_2O_4$** | MOCVD | 100 | Sapphire | 80 | 5.1 | $7.6\times10^{-4}$ | $1.6\times10^{15}$ | 7-10 | [90] |

**Table 4:** Binary oxide materials as *p*-type TCO and their respective optoelectronic parameters.

| Material | Deposition technique used | Thickness (nm) | Substrate used | Average optical Transmittance (%) | Optical band gap (eV) | Electrical conductivity (S/cm) | Carrier conc ($/cm^3$) | Hall Mobility ($cm^2/V\text{-}s$) | Reference |
|---|---|---|---|---|---|---|---|---|---|
| **NiO** | Magnetron Sputtering | 110 | glass | 40 | 3.8 | 7.2 | $1.3\times10^{19}$ | - | [5] |
| **NiO** | RFMS | 100 | glass | 69 | - | 9.09 | $8.8\times10^{18}$ | 6.12 | [96] |
| **K:NiO** | PLD | 100 | glass | 60 | 3.73-3.88 | 4.25 | $7.18\times10^{19}$ | 0.37 | [97] |
| **Cu-NiO** | Reactive Sputtering | 500 | Silicon | - | 2.41 | 8.34 | $4.22\times10^{20}$ | 1 | [98] |
| **$Sn_xNi_yO_z$** | RFMS | 100 | InP | - | 2.9 | 0.071 | $2.04\times10^{17}$ | 7.7 | [101] |
| **Mn:NiO** | spray pyrolysis | 200 | glass | 84 | 3.61 | 1.19 | - | - | [102] |
| **Zn:NiO** | spray pyrolysis | 200 | glass | 73 | 3.67 | 3.79 | - | - | [102] |
| **ZnO** | PLD | 700-1400 | Fused Silica | 80 | - | 0.5 | $5.6\times10^{18}$ | 0.1 | [107] |
| **SnO** | PLD | 110 | YSZ | - | 2.7 | - | $2.5\times10^{17}$ | 2.4 | [121] |
| **$SnO_x$** | RFMS | 200 | glass | 85 | 2.8 | 0.1-0.01 | $10^{17}$-$10^{18}$ | 4.8 | [123] |
| **SnO** | RFMS | 200-300 | Soda lime glass | 57 | 2.7-2.9 | 1.16 | $45.7\times10^{17}$ | 1.6 | [124] |
| **Ga-SnO** | RFMS | 200-300 | Soda lime glass | 65 | 2.7-2.9 | 3.71 | $62.4\times10^{17}$ | 3.6 | [124] |
| **Na-SnO** | RFMS | 200-300 | Soda lime glass | 50 | 2.7-2.9 | 90.9 | $469\times10^{17}$ | 11.3 | [124] |
| **$Ga:SnO_2$** | DC magnetron sputtering | 200 | Quartz | 85 | 3.8 | 0.021 | $8.84\times10^{18}$ | 0.0146 | [129] |
| **Eu-doped $SnO_2$** | spray pyrolysis | 700 | Si (100) and glass | 70-80 | 3.68-3.73 | 0.02-0.06 | $9.12\times10^{19}$ | 57.1-2.71 | [130] |
| **$Cu_2O$** | PLD | 100 | MgO | - | - | - | $10^{20}$ | 90 | [140] |
| **$Cu_2O$** | Thermal Evaporation | 223 | ITO and glass | - | 2.0 | $2.74\times10^{-3}$ | $2.8\times10^{15}$ | 6.1 | [142] |

**Table 5:** Chalcogen-based materials as *p*-type TCO and their respective optoelectronic parameters.

| Material | Deposition technique used | Thickness (nm) | Substrate used | Average optical Transmittance (%) | Optical band gap (eV) | Electrical conductivity (S/cm) | Carrier conc (/cm$^3$) | Hall Mobility (cm$^2$/V-s) | Reference |
|---|---|---|---|---|---|---|---|---|---|
| **$La_{1-x}Sr_xCuOS$** | RF Sputtering | 130 | Silica glass | >70 | 3.1 | 0.26 | - | - | [53] |
| **$LaCuOS_{1-x}Se_x$** | Reactive Solid phase | 150 | MgO | - | - | 24 | ~$10^{18}$ | 8 | [54] |
| **Mg-LaCuOSe** | Reactive Solid phase | 40 | MgO | 70 | 2.8 | 140 | $2.2\times10^{20}$ | 3.5 | [56] |
| **$(La_{1-x}Sr_xO)CuS$** | RF Sputtering | 150 | $SiO_2$ | >60 | - | 20 | - | - | [56] |
| **$La_{1-x}Y_xCuOS$** | Spin coating | 250 | Quartz | ~70 | 3 | 89.3 | $6.6\times10^{20}$ | 0.85 | [147] |
| **$La_{1-x}Sr_xCuOS$** | RF Sputtering | 130 | Silica glass | >70 | 3.1 | 0.26 | - | - | [53] |

**Table 6:** Cr-based materials as *p*-type TCO and their respective optoelectronic parameters.

| Material | Deposition technique used | Thickness (nm) | Substrate used | Average optical Transmittance (%) | Optical band gap (eV) | Electrical conductivity (S/cm) | Carrier conc (/cm$^3$) | Hall Mobility (cm$^2$/V-s) | FOM (µS) | Reference |
|---|---|---|---|---|---|---|---|---|---|---|
| **Mg,N:$Cr_2O_3$** | Spray pyrolysis | 150 | Glass | 10-65 | 3.1 | 0.06 | $10^{19}$ | 0.1 | 7.2 | [152] |
| **Ni:$Cr_2O_3$** | PLD | 150 | Sapphire | 40 | - | 28 | - | - | 110 | [154] |
| **$Cr_2O_3$** | PLD | 75 | Sapphire | 80 | 3.64 | 0.008 | $1.93\times10^{18}$ | 0.1 | 0.29 | [155] |
| **Mg:$Cr_2O_3$** | PLD | 75 | Sapphire | 70 | 3.54 | 0.4 | $2.04\times10^{18}$ | 0.4 | 8.16 | [155] |
| **Al:$Cr_2O_3$** | PLD | 75 | Sapphire | 81 | 3.59 | 0.009 | $1.90\times10^{18}$ | 0.11 | 0.33 | [155] |
| **Ni:$Cr_2O_3$** | PLD | 75 | Sapphire | 78 | 3.49 | 1.17 | $7.24\times10^{18}$ | 2 | 34.62 | [156] |
| **Sr-$LaCrO_3$** | MBE | 50 | STO | 42.3 | 1.9 | 54 | $7.5\times10^{21}$ | 0.04 | 3050 | [158] |
| **$La_{0.75}Sr_{0.25}CrO_3$** | Spin coating | 45 | STO and LAO | 67 | - | 1 | - | - | - | [159] |
| **Mg,N:$Cr_2O_3$** | Spray pyrolysis | 150 | Glass | 10-65 | 3.1 | 0.06 | $10^{19}$ | 0.1 | 7.2 | [152] |

**Table 7:** Some other oxide materials as *p*-type TCO and their respective optoelectronic parameters.

| Material | Deposition technique used | Thickness (nm) | Substrate used | Average optical Transmittance (%) | Optical band gap (eV) | Electrical conductivity (S/cm) | Carrier conc (/cm$^3$) | Hall Mobility (cm$^2$/V-s) | FOM (µS) | Reference |
|---|---|---|---|---|---|---|---|---|---|---|
| **$Ca_3Co_4O_9$** | Spin coating | 100 | Quartz | 67 | 3 | 18 | - | - | 151 | [163] |
| **$Bi_2Sr_2Co_2O_y$** | PLD | 50 | LAO | 51 | 3.14 | 181.8 | $5,17\times10^{20}$ | 2.2 | 1350 | [166] |
| **$In_2O_3$-$Ag_2O$** | RF Sputtering | 300 | Glass | 20 | - | 113.6 | $4.2\times10^{19}$ | 17 | - | [167] |
| **$In_2O_3$-$MoO_3$** | Spin coating | 80 | Glass | 80-88 | 3.8 | 400 | $2.1\times10^{20}$ | 11.9 | - | [168] |
| **$V_2O_3$** | PLD | 7-58 | Sapphire | 40-78 | 3.6 | 400-2122 | $>10^{22}$ | 0.14-0.24 | 12992 | [169] |
| **$V_2O_3$** | Spray pyrolysis | 60 | Sapphire | 32 | - | 1079 | $2.86\times10^{22}$ | 0.24 | 5699 | [170] |
| **ZnMgAlO** | Spray Pyrolysis | 200 | Borosilicate glass | 87.1 | 3.34 | 1 | $\sim10^{18}$ | 3.5 | - | [171] |
| **Cu-Mg $(OH)_2$** | Electro-chemical | 1000-2000 | ITO coated glass | 90 | 4 | $1\times10^{-5}$ | - | - | - | [172] |
| **$La_{1-x}Sr_xVO_3$** | PLD | 12 | LSAT | 53.9–70.1 | 3.5 | 742.3–872 | $1.72\times10^{21}$ | 2.1 | 2507-6776 | [174] |